\documentclass[aps,twocolumn,showpacs,groupedaddress, nofootinbib]{revtex4}  
\usepackage{graphicx} 
\usepackage{dcolumn}   
\usepackage{bm}        
\usepackage{amssymb}   
\usepackage{acronym}   
\usepackage{float}
\usepackage{xcolor}
\usepackage{hyperref}
\usepackage{latexsym}
\usepackage{url}
\usepackage{multirow}
\usepackage[flushleft]{threeparttable}
\usepackage{natbib}
\usepackage{subfigure}
\usepackage{amsmath}
\begin{document}

\widetext


\title{Estimate of the Detectability of the Circular Polarisation Signature of Supernova Gravitational Waves Using the Stokes Parameters}

\author{Man Leong Chan$^{1}$ \& Kazuhiro Hayama$^{1}$}
\affiliation{
$^1$Department of Applied Physics, Fukuoka University, Nanakuma 8-19-1, Fukuoka 814-0180, Japan}
\begin{abstract}
The circular polarisation of gravitational waves from core collapse supernovae has been proposed as a probe to investigate the rotation and physical features inside the core of the supernovae.
However, it is still unclear as to how detectable the circular polarisation of gravitational waves will be.
We developed an algorithm referred to as the Stokes Circular Polarisation algorithm for the computation of the Stokes parameters that works with the burst search pipeline coherent WaveBurst.
Employing the waveform SFHx and the algorithm, we estimate the detectability of the circular polarisation signatures ($V$ mode of the Stokes parameters) 
for sources across the sky at three different distances $2$, $5$, and $10$ kpc, 
for a network of gravitational wave detectors consisted of advanced LIGO, advanced VIRGO and KAGRA.
Using the Bayes factor, we found that for $2$ kpc and $5$ kpc, the majority of the sources ($99.9\%$ and $58.2\%$ respectively) will have their $V$ mode detectable,
while for $10$ kpc, no significant $V$ mode is detectable.
In addition, the significance of the $V$ mode signature are consistent with the recoverability of the two polarisations of gravitational waves
with respect to the network.
\end{abstract}
\maketitle
\acrodef{GW}[GW]{Gravitational wave}
\acrodef{BNS}[BNS]{binary neutron star}
\acrodef{BBH}[BBH]{binary black hole}
\acrodef{NSBH}[NSBH]{neutron star black hole}
\acrodef{EM}[EM]{electromagnetic}
\acrodef{CBC}[CBC]{compact binary coalescence}
\acrodef{CNN}[CNN]{Convolutional neural network}
\acrodef{CCSN}[CCSN]{core collapse supernova}
\acrodefplural{CCSN}[CCSNe]{core collapse supernovae}
\acrodef{ROC}[ROC]{receiver operator characteristic}
\acrodef{TAP}[TAP]{true alarm probability}
\acrodef{FAP}[FAP]{false alarm probability}
\acrodef{aLIGO}[aLIGO]{advanced LIGO}
\acrodef{AdVirgo}[AdVirgo]{advanced VIRGO}
\acrodef{SASI}[SASI]{standing accretion shock instability}
\acrodef{CWB}[cWB]{coherent WaveBurst}
\acrodef{SNR}[SNR]{signal to noise ratio}
\acrodef{SCP}[SCP]{Stokes Circular Polarisation}
\section{Introduction}
\ac{GW} astronomy has been rapidly developing since the first direct detections from the O1 and O2 observations of LIGO and VIRGO~
\cite{abbott2016observation, abbott2016gw151226, abbott2017gw170608, abbott2017gw170814, abbott2017gw170817, abbott2017gravitational, abbott2017multi, abbott2019gwtc}.
As the sensitivities of the interferometric \ac{GW} detectors improve, 
more detections of transient \ac{GW} events are anticipated~\cite{abbott2018prospects}.
Although at present all the detections have originated from compact binary coalescences,
\acp{CCSN} are among the sources of \acp{GW} expected to be observable with the second generation detectors 
such as \ac{aLIGO}, \ac{AdVirgo} and KAGRA~\cite{aasi2015advanced, acernese2014advanced, aso2013interferometer, gossan2016observing, abbott2016first}.

As a massive star ($10$ - $100~\text{M}_\odot$ at zeros-age main sequence) reaches the final stage of 
its stellar life after all its stellar fuel has been combusted via nuclear reaction, 
core collapse is expected to ensue if the mass of the core is larger than the effective Chandrasekhar mass~\cite{baron1990effect, bethe1990supernova}.
The core collapse will continue until its density is comparable to that of nuclear matter.
The inner core will bounce back as the nuclear equation of state stiffens by the strong nuclear force.
A shock wave will then be formed and sent through the infalling matter.
By losing energy to the dissociation of the iron nuclei and to neutrino cooling, the shock wave will stall, which
will somehow have to be revived if the star is to become a supernova~\cite{o2011black}.
Although this scenario is supported by the observations of the \ac{CCSN} SN1987A \cite{sato1987analysis} in 1987, 
how exactly the shock wave is revived is still unclear to astronomers and has remained the subject of intense study for decades~\cite{janka2012explosion}.

Currently, two most popular theories, the neutrino-driven mechanism \cite{bethe1985revival, bethe1990supernova} and the magnetorotational mechanism,
for reviving the shock wave in the inner core, have been put forward.
For stars with rapid core spin and a strong magnetic field,
the magnetorotational mechanism may be the active mechanism~\cite{leblanc1970numerical, burrows2007simulations, 
takiwaki2009special, moiseenko2006magnetorotational,mosta2014magnetorotational} 
(such a requirement may not be absolutely necessary, see e.g. \cite{obergaulinger2019magnetorotational}). 
The magnetic field and the core spin together may
produce an outflow that could possibly cause some of the most energetic \acp{CCSN} observed and may be able to explain the extreme hypernovae and the observed long gamma-ray
bursts~\cite{woosley2006progenitor, yoon2005evolution, de2013rotation, obergaulinger2017protomagnetar}.
On the other hand, the neutrino-driven mechanism\cite{bethe1985revival, bethe1990supernova} theorises that the revival of the shock wave is achieved by $\sim5\%$
of the outgoing neutrino energy stored below the shock, which causes turbulence and increases thermal pressure.
Convection and the \ac{SASI}\cite{blondin2003stability}, observed  in supernova simulations, may also be required to produce a \ac{CCSN} via the neutrino mechanism.  

Related to the mechanisms, one important property of massive stars is their rotation profiles, such as the rate of the rotation and the differential rotation.  
Rotation itself is also a parameter important to the chemical yields, stellar evolution as well as the final stage of its life~\cite{langer2012presupernova}.
However,  although stars are generally known to be rotators \cite{ramirez2014rotational}, the rates at which stars rotate 
in their evolution history and just prior to collapse are still unknown.
In part, the rotation of a star may depend on the presence of the magnetic braking of the rotation. 
It is estimated that periods of a few seconds are possible if the braking is not present, 
while the periods can be up to ten times longer with the presence of angular momentum transfer via magnetic
fields~\cite{maeder2012rotating}. 

Methods for investigating the rotation of the cores have been proposed in the literature.
For example, signature of the angular momentum distribution was once suggested to be 
imprinted in the sign of the second largest peak in the \acp{GW} emitted after core bounce~\cite{hayama2008determination}.
In addition, \cite{abdikamalov2014measuring} proposed the use of a waveform template 
bank for signals from sources with rapid rotations, as well as Bayesian model selection~\cite{abdikamalov2014measuring}.
In recent years, the circular polarisation of \acp{GW} has been proposed as a probe
to investigate the rotation of the core prior to collapse~\cite{hayama2016circular, hayama2018circular}. 
It was pointed out that rapidly rotating cores of massive stars can cause the formation of accretion flows that have non-axisymmetric, 
spiral pattern in the post-shock (for example, see. e.g. \cite{kuroda2014gravitational, seto2007measuring}). 
The core rotation might also reflect itself as a signature of circular polarisation in the emitted \acp{GW} at frequency
twice that of the rotation~\cite{hayama2016circular}.

Other than rotation, the circular polarisation of a \ac{GW} may also help understand the physical features deep in 
the core of a supernova.
For instance, it has been recognised that the circular polarisation signature of the \acp{GW} from \acp{CCSN}
may contain information on the \ac{SASI} activity ~\cite{kuroda2016new, andresen2017gravitational, hayama2018circular}
and the ramp-up $g$-mode of the proto-neutron star oscillation and can therefore be used as a probe of these features.
The circular polarisation of \acp{GW} may also show the evolution of the asymmetry between the right-handed and left-handed mode (defined in section \ref{sec:stokes}) 
over time and frequency~\cite{hayama2018circular}.

However, the detection of a \ac{GW} signal depends on the combined antenna pattern of \ac{GW} detectors to the source location,
while the recoverability of the circular polarisation signature of a \ac{GW} relies on both the sensitivities of the \ac{GW} detectors to the two polarisations of the signal. 
Therefore, one question that can be asked is will we be able to recover the circular polarisation signatures of the \acp{GW} 
from \acp{CCSN} if such signals are detected? 
In \cite{hayama2018circular}, the authors tried to answer such a question. 
The answer turned out to be positive. The analysis, however, was based on only one example.
This means that the conclusion itself may not be entirely representative.

In this work, we extend the method and the work presented in~\cite{hayama2016circular, hayama2018circular}, 
and develop an algorithm that computes the Stokes parameters in the time-frequency domain.
The algorithm works with the detection pipeline \ac{CWB}~\cite{2016PhRvD..93d2004K}, which 
is one of the main detection pipelines employed in LIGO and VIRGO and was among the first pipelines to achieve the first direct detection of \acp{GW}~\cite{abbott2016observing}.
Using the simulated waveform SFHx from~\cite{kuroda2016new} as an example and the developed algorithm, we
try to answer the question in a more general manner by performing simulations of sources across the sky.
This paper is structured as follows: in section \ref{sec:stokes}, we will present a brief explanation of the Stokes parameters.
Section \ref{sec:al} will then be devoted to the algorithm developed for the computation of the parameters. In \ref{sec:vobsered},
the details of the simulations are given. The results and a discussion are presented in section \ref{sec:disuss}, which is followed by a conclusion in section \ref{sec:conclusion}.



\section{The Stokes parameters}\label{sec:stokes}
The Stokes parameters \cite{seto2007measuring} are a set of physical values that can be used to describe the polarisation status of \acp{GW}. 
The mathematical definition of the Stokes parameters is given by,
\begin{align}\label{eq:stokesp}
 \langle h_R(f, \hat{n}) h_R(f', \hat{n}')^{*}\rangle = \frac{1}{4\pi}\delta_{D}^2(\hat{n} - \hat{n}')\delta_{D}^2(f - f') \times \nonumber\\
 (I(f, f', \hat{n}, \hat{n}') + V(f, f', \hat{n}, \hat{n}')); \nonumber\\
 \nonumber\\ 
 \langle h_L(f, \hat{n}) h_R(f', \hat{n}')^{*}\rangle = \frac{1}{4\pi}\delta_{D}^2(\hat{n} - \hat{n}')\delta_{D}^2(f - f') \times \nonumber\\
 (Q(f, f', \hat{n}, \hat{n}') - iU(f, f', \hat{n}, \hat{n}')); \nonumber\\ 
 \nonumber\\ 
 \langle h_R(f, \hat{n}) h_L(f', \hat{n}')^{*}\rangle = \frac{1}{4\pi}\delta_{D}^2(\hat{n} - \hat{n}')\delta_{D}^2(f - f') \times \nonumber\\
 Q(f, f', \hat{n}, \hat{n}') + iU(f, f', \hat{n}, \hat{n}')); \nonumber\\ 
 \nonumber\\ 
 \langle h_L(f, \hat{n}) h_L(f', \hat{n}')^{*}\rangle = \frac{1}{4\pi}\delta_{D}^2(\hat{n} - \hat{n}')\delta_{D}^2(f - f') \times \nonumber\\
 I(f, f',  \hat{n}, \hat{n}') - V(f, f', \hat{n}, \hat{n}'));
\end{align} 
where $f$ is frequency, $\hat{n}$ the unit vector in the propagation direction, and $\langle \rangle$ represents the ensemble average.
In the above equation, $h_R$ and $h_L$ are  the right-handed and left-handed mode of \ac{GW}, as given by
\begin{align}\label{eq:rhlh}
 h_R &\equiv \frac{(h_{+} - \text{i}h_{\times})}{\sqrt{2}},\nonumber \\
 h_L &\equiv \frac{(h_{+} + \text{i}h_{\times})}{\sqrt{2}},
\end{align}
where the terms $h_{+}$ and $h_{\times}$ are the two polarisations of \ac{GW}.
$I, Q, U$ and $V$ are the full set of the Stokes parameters.
They describe different properties of a \ac{GW}. 
For instance, the parameter $I$ represents the total amplitudes of the right-handed and left-handed mode,
$Q$ and $U$ the linear polarisation status. 
In particular, the parameter $V$  describes the circular polarisation.
Since we are interested in the circular polarisation of \acp{GW}, 
we will focus on the $V$ parameter, which we will refer to as the $V$ mode in this paper.
With some algebraic manipulations, it can be shown from Eq. \ref{eq:stokesp} that the $V$ mode can be written as,
\begin{align}\label{eq:stokesV}
 \delta_{D}^2(\hat{n} - \hat{n}')\delta_{D}^2(f - f')V = 2\pi \langle h_R(f, \hat{n}) h_R(f', \hat{n}')^{*}\rangle  \nonumber \\
 -  \langle h_L(f, \hat{n}) h_L(f', \hat{n}')^{*}\rangle,
\end{align}
in other words, the $V$ mode also describes the amplitude asymmetries between the right-handed and left-handed mode.
\section{Algorithm}\label{sec:al}
\begin{figure}
\includegraphics[width=0.48\textwidth]{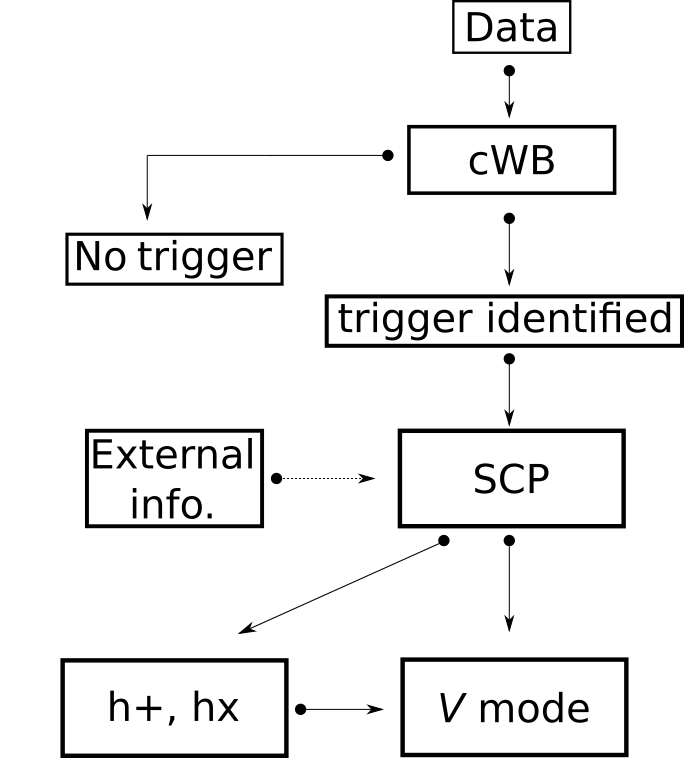}
\caption{An flow chart showing the work flow of the algorithm for the $V$ mode computation.
It works by first extracting the whitened time series from the pipeline \ac{CWB} if there is a trigger identified by the pipeline. 
Using either the estimates on the arrival time of the event and the sky location from the \ac{CWB} or 
external information such as that from electromagnetic observations or neutrino observations, 
the algorithm will then recover the two polarisations $h_+$ and $h_{\times}$.
From the two polarisations, the algorithm will then compute $V$ mode of the trigger.
The dashed line connecting External info. and \ac{SCP} indicates that the algorithm may or may not use the information from external sources.
\label{fig:flowchart}}
\end{figure}

As mentioned in the introduction, we develop an algorithm for the computation of the $V$ mode of \acp{GW} in the time-frequency domain.
The algorithm is illustrated in Fig. \ref{fig:flowchart}. For the remaining of this paper, the algorithm will be referred to as the \ac{SCP} algorithm.
As shown in the flow chart, if a trigger is identified by \ac{CWB} and the \ac{SCP} algorithm is called, 
it will first input the corresponding whitened time series at time $t$ from \ac{CWB} as $\bold{d}(t)$, given by
\begin{equation}
 \bold{d}(t) = \bold{F}_{+}h_{+}(t) + \bold{F}_{\times}h_{\times}(t),
\end{equation}
where the antenna pattern of the detectors is denoted by $\bold{F}_+$ and $\bold{F}_{\times}$.
Since the $V$ mode describes the asymmetry of the right-handed and left-handed mode of a \ac{GW}, which are related to the two polarisations $h_{+}$ and $h_{\times}$
by Eqs. \ref{eq:rhlh} and \ref{eq:stokesV},
the first step is to recover $h_+$ and $h_\times$ from $\bold{d}$.
This is achieved in the \ac{SCP} algorithm by using the following equation~\cite{di2018estimation},
\begin{equation}\label{eq:reconstruction}
\begin{cases} 
h^{r}_+(t) = \frac{(\bold{F}_+\cdot \bold{d}(\bold{t}))(\lvert\bold{F}_{\times}\rvert^{2}) - 
(\bold{F}_{\times}\cdot \bold{d}(\bold{t}))(\bold{F}_{+} \cdot \bold{F}_{\times})}
{\lvert\bold{F}_{+}\rvert^{2}\lvert\bold{F}_{\times}\rvert^{2} - (\bold{F}_{+} \cdot \bold{F}_{\times})^2},\\ 
h^{r}_\times(t) = \frac{(\bold{F}_{\times}\cdot \bold{d}(\bold{t}))(\lvert\bold{F}_{+}\rvert^{2}) - 
(\bold{F}_{+}\cdot \bold{d}(\bold{t}))(\bold{F}_{+} \cdot \bold{F}_{\times})}
{\lvert\bold{F}_{+}\rvert^{2}\lvert\bold{F}_{\times}\rvert^{2} - (\bold{F}_{+} \cdot \bold{F}_{\times})^2},\\
\end{cases} 
\end{equation}
where $h^{r}_+(t)$ and $h^{r}_\times(t) $ are the reconstructed polarisation components and the superscript $r$ stands for reconstruction. 
To use Eqs. \ref{eq:reconstruction}, the \ac{SCP} algorithm can either employ the estimates of the sky location and the arrival time of the signal from the \ac{CWB},
or from electromagnetic and/or neutrino observations in situations where they are available.  
The next step is to reconstruct the $V$ mode of the signal using $h^{r}_+(t)$ and $h^{r}_\times(t) $, which will be explained in section \ref{sec:rev}.

\subsection{Reconstruction of $V$ mode}\label{sec:rev}
The easiest way to achieve the reconstruction of the $V$ mode of a signal in the time-frequency domain is to substitute $h^{r}_+(t)$ and $h^{r}_\times(t) $ for 
in $h_{+}$ and $h_{\times}$ respectively in Eq. \ref{eq:rhlh}. 
The $V$ mode can then be computed  using Eq. \ref{eq:stokesV}, which we denote by $\bold{V}_\text{o}$.

However, for a random trigger, $\bold{V}_\text{o}$ computed in such a way will not be entirely free of noise.
To determine the $V$ mode of a trigger, we further take the following procedure:
first, $N$ whitened time series, $\bold{d}^{\text{n}}_{k}$, of the same duration as the trigger, adjacent to the time of the event as estimated by \ac{CWB}, are taken.
The superscript $\text{n}$ indicates that these time series contain only noise, and the subscript $k$ ranges from $1$ to $N$.
The value $N$ is an arbitrary number chosen before the calculation.
We then compute the $V$ mode for each of $\bold{d}^{\text{n}}_{k}$ by substituting $\bold{d}^{\text{n}}_{k}$ for $\bold{d}$ in Eq. \ref{eq:reconstruction}, 
while using the same values of $\bold{F}_+$ and $\bold{F}_{\times}$
and the arrival times as those for $\bold{V}_\text{o}$.
The $V$ modes obtained are denoted by $\bold{V}_{k}^{\text{n}}$.
As $\bold{V}_{k}^{\text{n}}$ is computed in the time-frequency domain, each $\bold{V}_{k}^{\text{n}}$ contains $m$ time-frequency pixels, 
where $m$ is a number depending on the number of overlap and Fast Fourier transform window.
This means in total there will be $N\times m$ time-frequency pixels.

The final step is then to rank the  $N\times m$ pixels based on their absolute values. 
The value larger than a pre-selected fraction $\text{F}_o$ (e.g., $0.99$) of the pixels will be selected as a threshold $p_\text{thr}$.
For a pixel from $\bold{V}_\text{o}$ to be considered relevant to the trigger rather than random noise, its absolute value has to be larger than $p_\text{thr}$.
The collection of the pixels $> p_\text{thr}$ from $\bold{V}_\text{o}$ as well as their corresponding time and frequency 
is then the $V$ mode of the trigger, denoted by $\bold{V}_\text{t}$, where the subscript t stands for $>$ threshold.

\subsection{Significance of $V$ mode}\label{sec:sv}
Once  $\bold{V}_\text{t}$  of a trigger is reconstructed, it is necessary to establish its significance.
To do so, we employ the Bayes factor.
The Bayes factor is a ratio of the posteriors of two competing models or hypotheses, as defined by the following equation,
\begin{equation}\label{eq:BF}
 B_{\text{H}_1 / \text{H}_0} = \frac{p(\text{H}_1 |\bold{d})}{p(\text{H}_0|\bold{d})},
\end{equation}
where $p(\text{H}_0|\bold{d})$ 
is the posterior of the null hypothesis $\text{H}_0$ given $\bold{d}$, 
and $p(\text{H}_1|\bold{d})$ the posterior of the alternative hypothesis given $\bold{d}$.
The Bayes factor measures how much a hypothesis or a model is favoured by the data against another competing hypothesis or model.
If the value of $B_{\text{H}_1 / \text{H}_0}$ is larger than $1$, it means $\text{H}_1$ is favoured by the data, 
or the data is in favour of $\text{H}_0$ if the value is less than $1$. 
While Bayes factors is similar to the signal-to-noise ratio of a signal in the sense that they can both be used to estimate the significance of the presence of a signal,
they are not equal.

Since our purpose is to determine the presence or absence of the $V$ mode of a trigger,
$\text{H}_1$ is the hypothesis that a $V$ mode signature is detected, 
and $\text{H}_0$ no $V$ mode signature is detected.  
Using the Bayes' theorem and substituting $\bold{V}_\text{o}$ for $\bold{d}$, Eq. \ref{eq:BF} can be written as,
\begin{equation}\label{eq:BF2}
 B_{\text{H}_1 / \text{H}_0} = \frac{p(\text{H}_1)}{p(\text{H}_0)}\frac{p(\bold{V}_\text{o}|\text{H}_1)}{p(\bold{V}_\text{o}|\text{H}_0)}.
\end{equation}
In the above equation, $p(\text{H}_1)$ and $p(\text{H}_0)$ are the prior probabilities for $H_1$ and $H_0$ respectively, which we set to be equal.
The terms $p(\bold{V}_\text{o}|\text{H}_1)$ is the likelihood of $\bold{V}_\text{o}$ given $\text{H}_1$ and $p(\bold{V}_\text{o}|\text{H}_0)$ the likelihood of 
$\bold{V}_\text{o}$ given $\text{H}_0$.	
We approximate the likelihood function using a Gaussian distribution,
so the joint probability densities can be written as,
\begin{align}\label{eq:posterior}
 p(\bold{V}_\text{o}|\text{H}_1) &= \prod_{i}^{m} \frac{1}{\sqrt{2\pi} \sigma_i}\exp{-\frac{1}{2m}(\frac{V_{\text{o}i} - V_{\text{t}i} - V_{\text{M}i}}{\sigma_i} )^2} \\ \nonumber
 p(\bold{V}_\text{o}|\text{H}_0) &= \prod_{i}^{m} \frac{1}{\sqrt{2\pi} \sigma_i}\exp{-\frac{1}{2 m}(\frac{V_{\text{o}i}- V_{\text{M}i}}{\sigma_i})^2},
\end{align}
where $V_{\text{o}i}$, $V_{\text{t}i}$ and $V_{\text{M}i}$ are the $i$th pixel from 
$\bold{V}_{\text{o}}$, $\bold{V}_{\text{t}}$ and $\bold{V}_{\text{M}}$ respectively.
$\bold{V}_{\text{M}}$ and $\sigma$ are the mean and standard deviation of $\bold{V}_{k}^{\text{n}}$ respectively. 
In addition, we normalise the likelihood functions by taking the geometric mean of the functions over the number of pixels ($m$th root).
This is to prevent artificial change of the value of Bayes factor due to the change of overlaps and the Fourier transform window.
Finally, for our analysis, we will use the logarithm of the Bayes factor $\log B_{\text{H}_1 / \text{H}_0}$ for the remaining of this paper.

\begin{figure*}
     \begin{center}
        \subfigure[]{
            \label{fig:sincgauwave1}
            \includegraphics[width=0.48\textwidth]{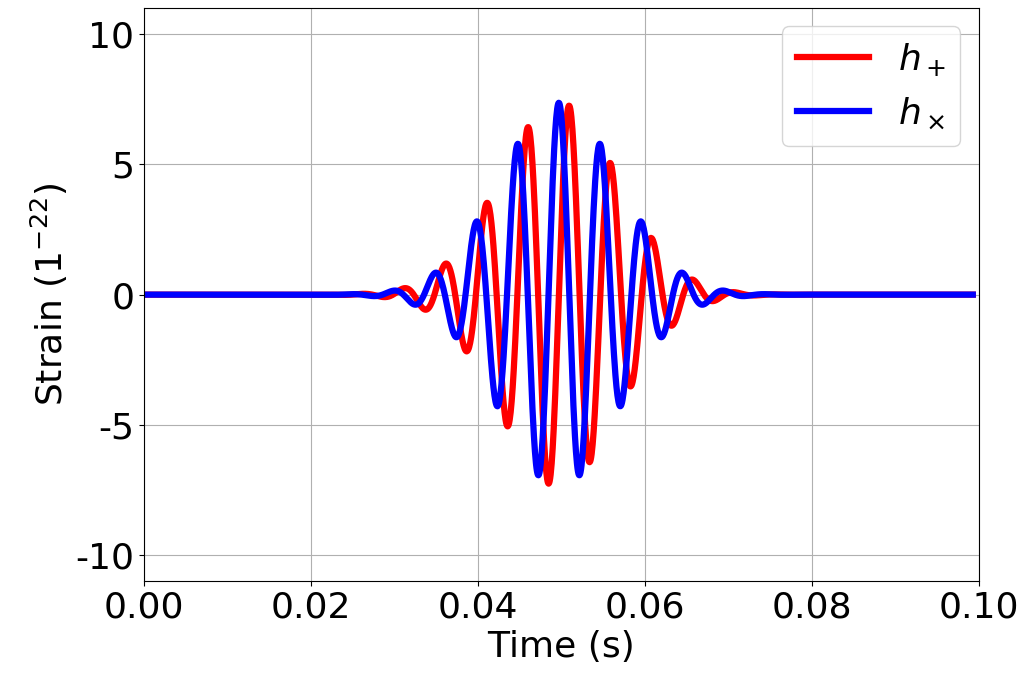}%
        }\quad
        \subfigure[]{
            \label{fig:sincgauwaveV1}
            \includegraphics[width=0.48\textwidth]{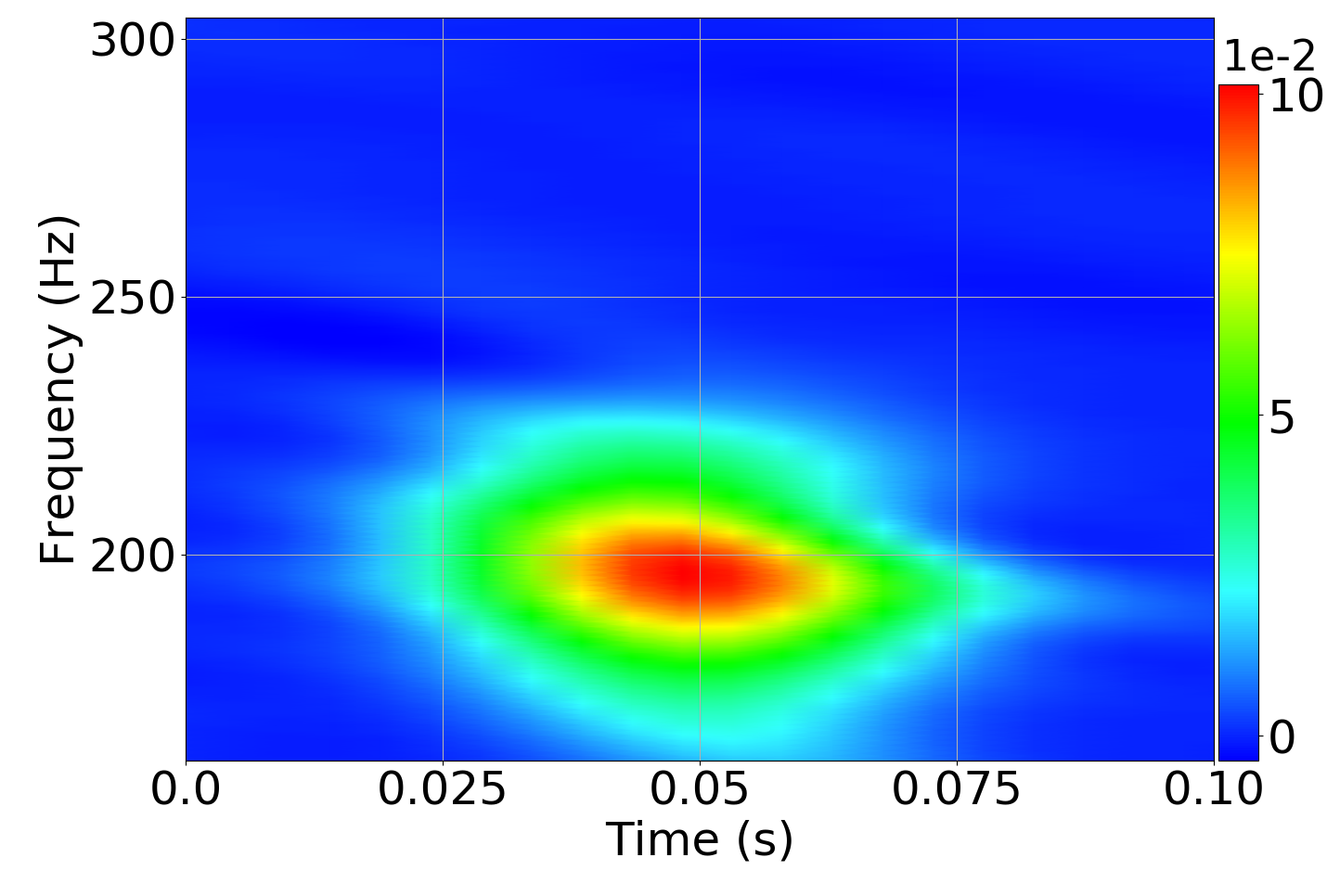}%
        }\quad
        \subfigure[]{
            \label{fig:sincgauwaveSNR1}
            \includegraphics[width=0.48\textwidth]{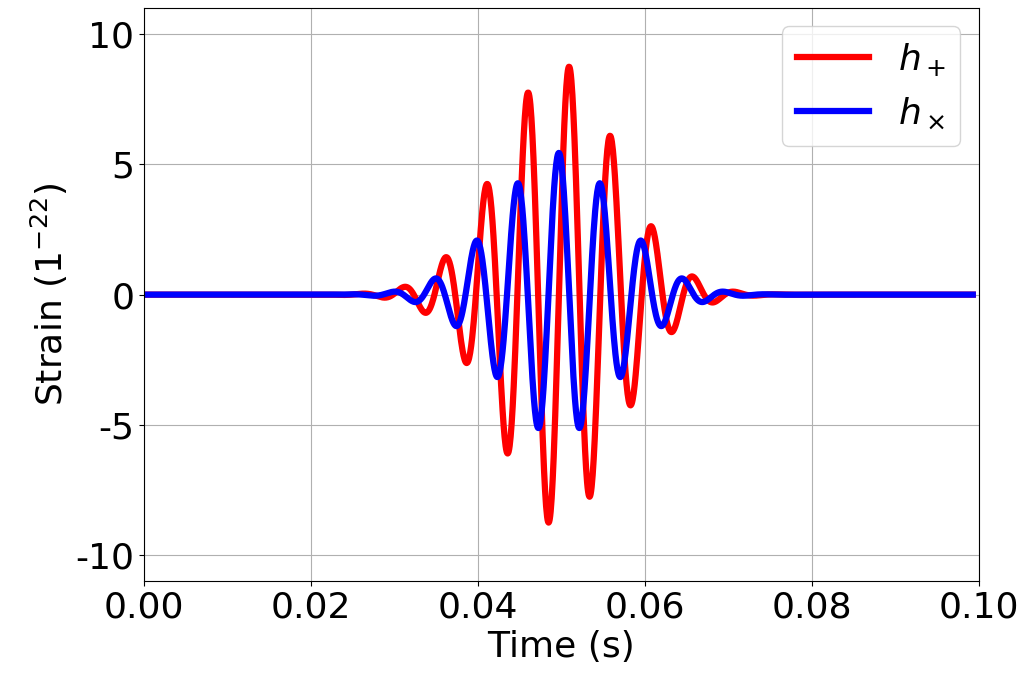}%
        }\quad
        \subfigure[]{
            \label{fig:sincgauwave2}
            \includegraphics[width=0.48\textwidth]{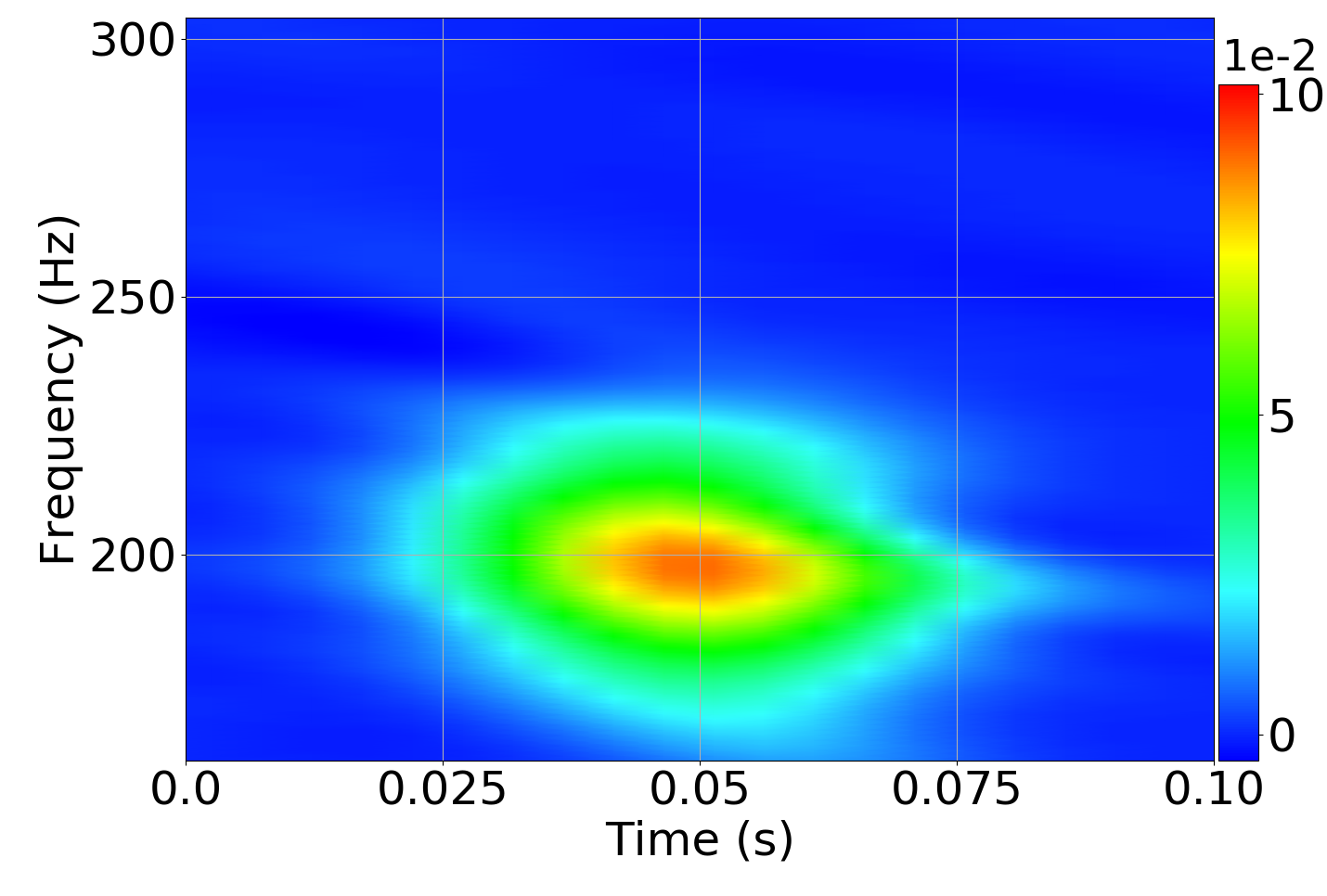}%
        }\quad
        \subfigure[]{
            \label{fig:sincgauwaveV2}
            \includegraphics[width=0.48\textwidth]{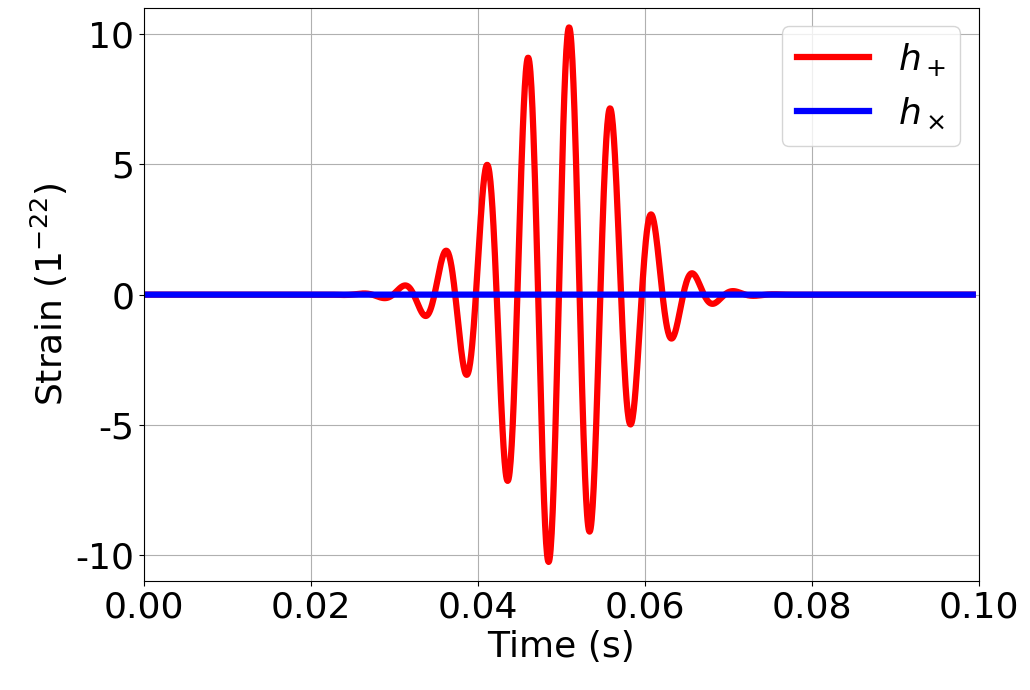}%
        }\quad
        \subfigure[]{
            \label{fig:sincgauwaveSNR2}
            \includegraphics[width=0.48\textwidth]{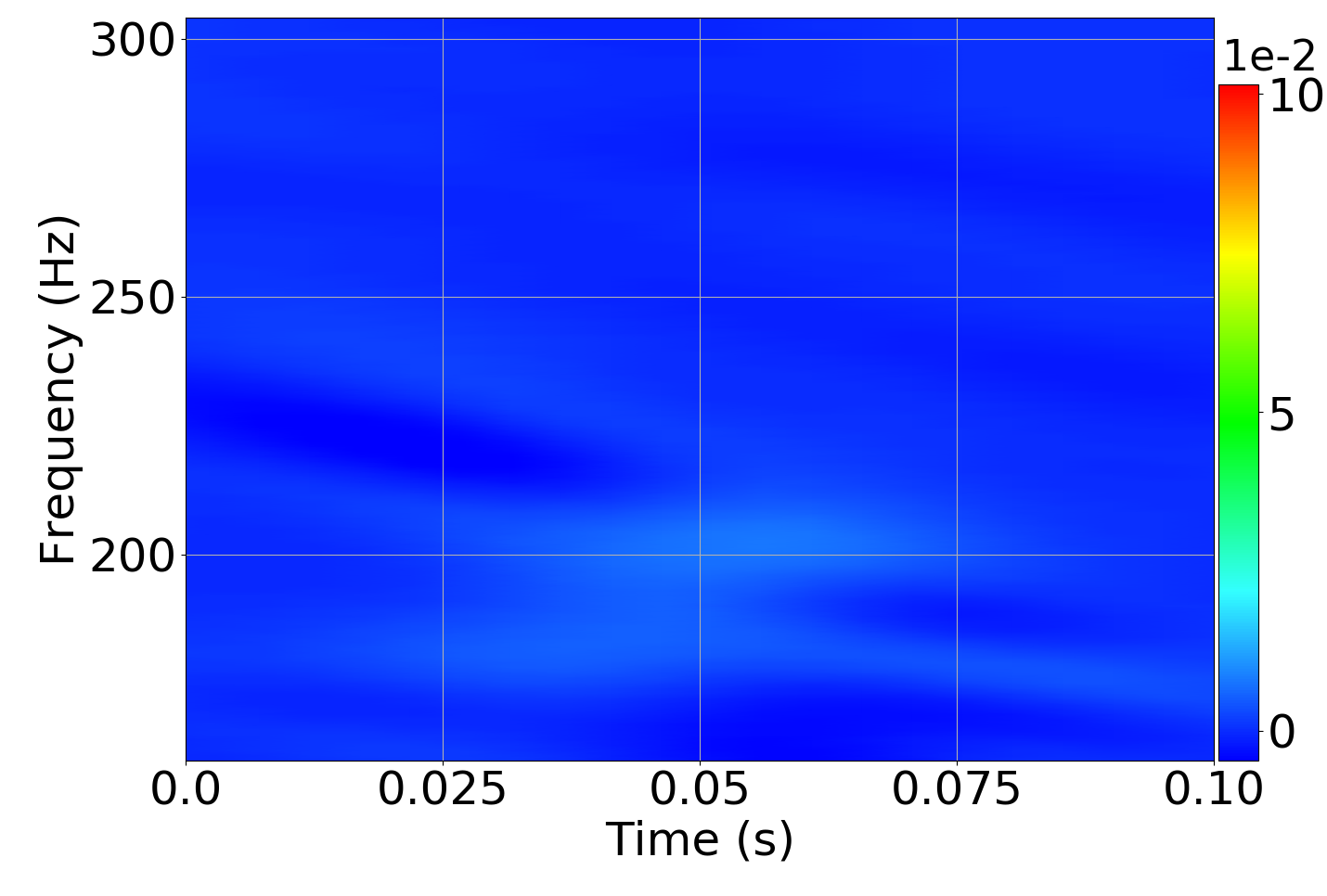}%
        }
    \end{center}
    \caption{Examples of sine-Gaussian waveforms and their $V$ mode presented in the time-frequency domain.
    From the top to the bottom, the left panels show the sine-Gaussian waveforms generated  at $\iota= 0^{\circ}, 70^{\circ}, 90^{\circ}$ respectively. 
    The amplitudes are normalised to  $h_\text{rss} = 8.3\times10^{-23}$.
    The sources are injected at  (longitude, latitude) = ($-90^\circ$, $30^\circ$).
    The panels on the right show the respective $V$ mode computed using the whitened time series from \ac{CWB}.
    \label{fig:simpleV}} 
\end{figure*}

As an example to show how the \ac{SCP} algorithm works, we test the algorithm against three cases where the signals are sine-Gaussian waves.
The sine-Gaussian waves are generated using the following equation,
\begin{align}\label{eq:singau}
 h_{+}(t) &= h\frac{1 + \text{cos}^2(\iota)}{2}\text{sin}(2\pi t f_0) e^{\frac{-2 \pi^2  f^2_0 t^2 }{Q^2}},\nonumber \\
 h_{\times}(t) &= h \text{cos}(\iota)\text{cos}(2\pi t f_0) e^{\frac{-2 \pi^2  f^2_0 t^2 }{Q^2}},
\end{align}
where $h$ is the amplitude, $Q$ quality factor, $f_0$  frequency, and $\iota$ the inclination angle.
For simplicity, we generate three distinct cases with three different values of $\iota$ (i.e., $0^{\circ}, 70^{\circ}, 90^{\circ}$ respectively), 
while keeping the values of $Q = 9$ and  $f_0 = 200$Hz for all three cases.
These three values of $\iota$ are chosen to represent different polarisation status, such as circular polarisation ($0^{\circ}$), 
elliptical polarisation ($\iota = 70^{\circ}$) and linear polarisation ($\iota = 90^{\circ}$), as seen from the observer.
For a fair comparison between these three cases, the sky locations of the sources are chosen to be the same at (longitude, latitude) = ($-90^\circ$ , $30^\circ$) and 
the values of $h$ are chosen  such that each sine-Gaussian wave has the same value of $h_\text{rss}$ (i.e. $8.3\times10^{-23}$).
The network consists of \ac{aLIGO} Hanford, \ac{aLIGO} Livingston, \ac{AdVirgo} and KAGRA. 
The noise is Gaussian noise generated using the power spectrum densities at their respective design sensitivity~\cite{abbott2018prospects, acernese2014advanced, akutsu2018kagra}.
For the settings in the \ac{SCP} algorithm, we choose $N$ to be 100 and $\text{F}_o$ to be $0.99$. 
The true values of arrival times and the antenna pattern are used.
The results are shown in Fig. \ref{fig:simpleV}.
From $\iota = 0$ to  $90^{\circ}$, a decrease in the magnitude of the $V$ mode can be seen.
This indicates that the magnitude of the $V$ mode captures the extent to which the signal is circularly polarised.
For the sine-Gaussian waves at $\iota=$ $0^{\circ}, 70^{\circ}, 90^{\circ}$, the $\log {B_{\text{H}_1 / \text{H}_0}}$ 
are  $67.2$, $55.0$ and $0.4$ respectively.  
If we consider $\text{H}_1$ is preferred when $\log B_{\text{H}_1 / \text{H}_0}>= 8$, 
then it can be seen that for the cases at $\iota=$ $0^{\circ}$ and  $70^{\circ}$, $\text{H}_1$ is preferred, 
while for $\iota=90^{\circ}$, $\text{H}_0$ is preferred, which are as expected.

\begin{figure}
     \begin{center}
        \subfigure[]{
            \label{fig:sfhxvo}
            \includegraphics[width=0.48\textwidth]{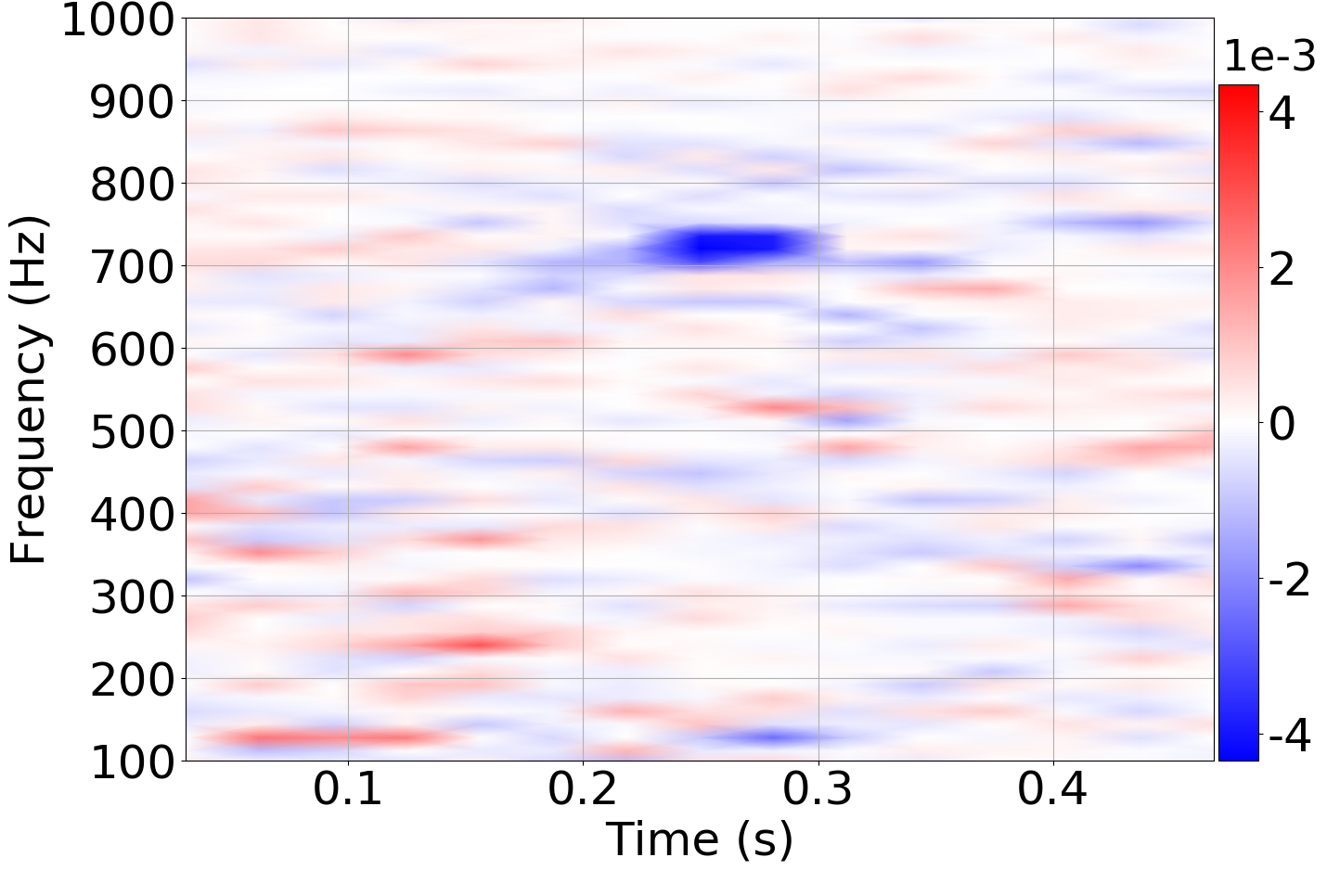}%
        }\quad
        \subfigure[]{
            \label{fig:sfhxvt}
            \includegraphics[width=0.48\textwidth]{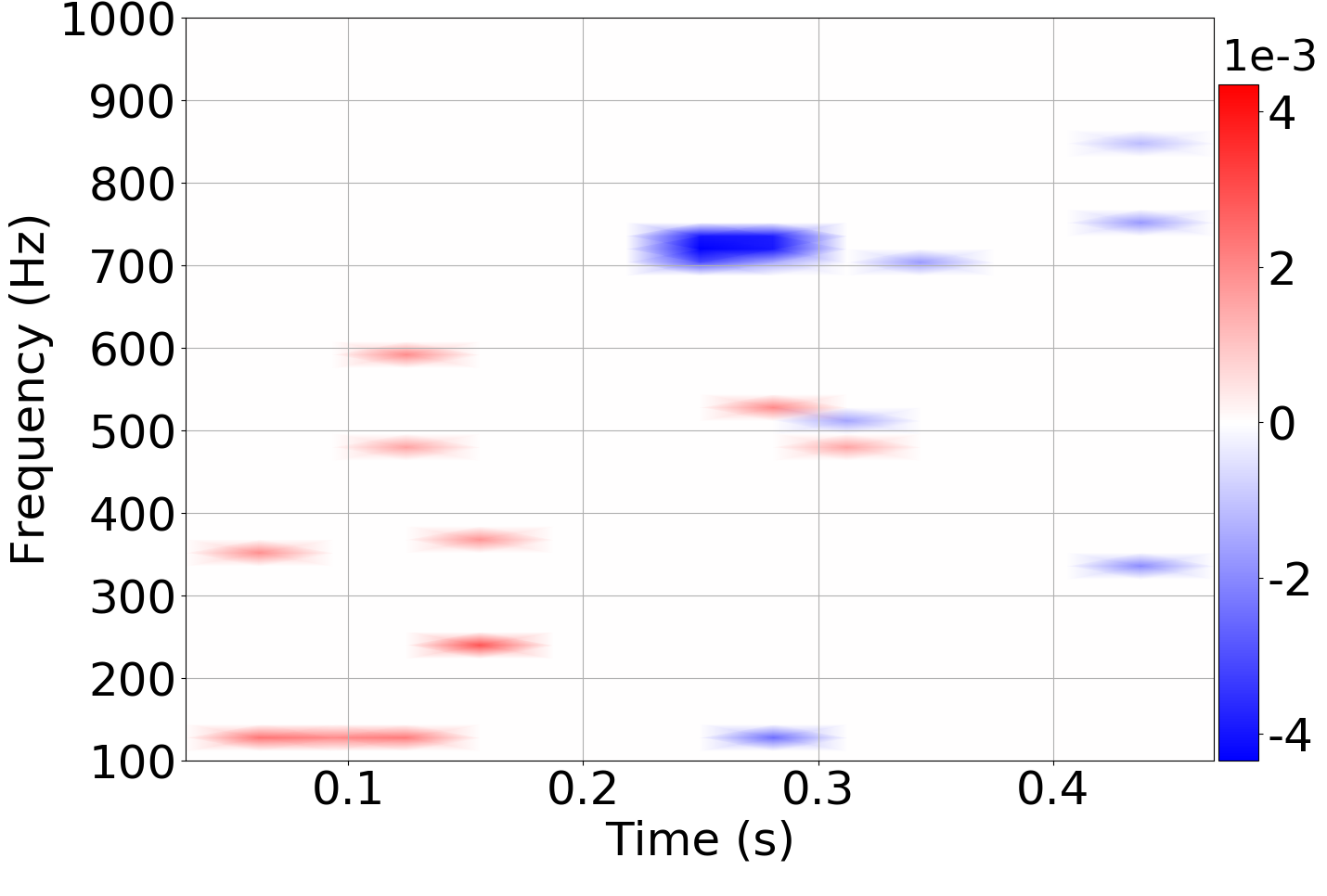}%
        }
    \end{center}
    \caption{The reconstruction of the $V$ mode of a trigger where the injected waveform is SFHx. 
    The upper panel shows the $\bold{V}_\text{o}$ and the lower panel shows $\bold{V}_\text{t}$ of the trigger.
    The source is located at (longitude, latitude) = ($-145^\circ$, $70^\circ$) and $10$ kpc from earth.
    The network of \ac{GW} detectors consists of \ac{aLIGO} Hanford, \ac{aLIGO} Livingston, \ac{AdVirgo} and KAGRA.
    \label{fig:reconstructedvmode}} 
\end{figure}

The above described approach can suppress most time-frequency pixels irrelevant to a trigger.
While this works well for strong signals, we also note that signals with pixels associated with weak amplitudes can potentially be ruled out, 
especially in situations where prior knowledge on the waveforms of the signals are not available.
As an example, we show the $\bold{V}_\text{o}$ and $\bold{V}_\text{t}$ of a trigger in Fig. \ref{fig:reconstructedvmode} 
where the injected  waveform is SFHx 
(see section \ref{sec:vobsered} for details of the waveform) 
and the source is located at (longitude, latitude) = ($-145^\circ$, $70^\circ$), and $10$ kpc from earth.
It can be seen that the peaks of $\bold{V}_\text{t}$ occur at times and frequencies that are generally consistent with those shown 
in Fig. \ref{fig:simpleV2}. 
However, the amplitudes of the signal are so weak that they do significantly deviate from noise and 
the value of $\log B_{\text{H}_1 / \text{H}_0}$ is $0.4$, 
indicating the difficulty of detecting and reconstructing the $V$ mode of a signal 
without any prior knowledge on the waveform.

\section{$V$ mode observability}\label{sec:vobsered}
\begin{figure*}
     \begin{center}
        \subfigure[]{
            \label{fig:domon}
            \includegraphics[width=0.48\textwidth]{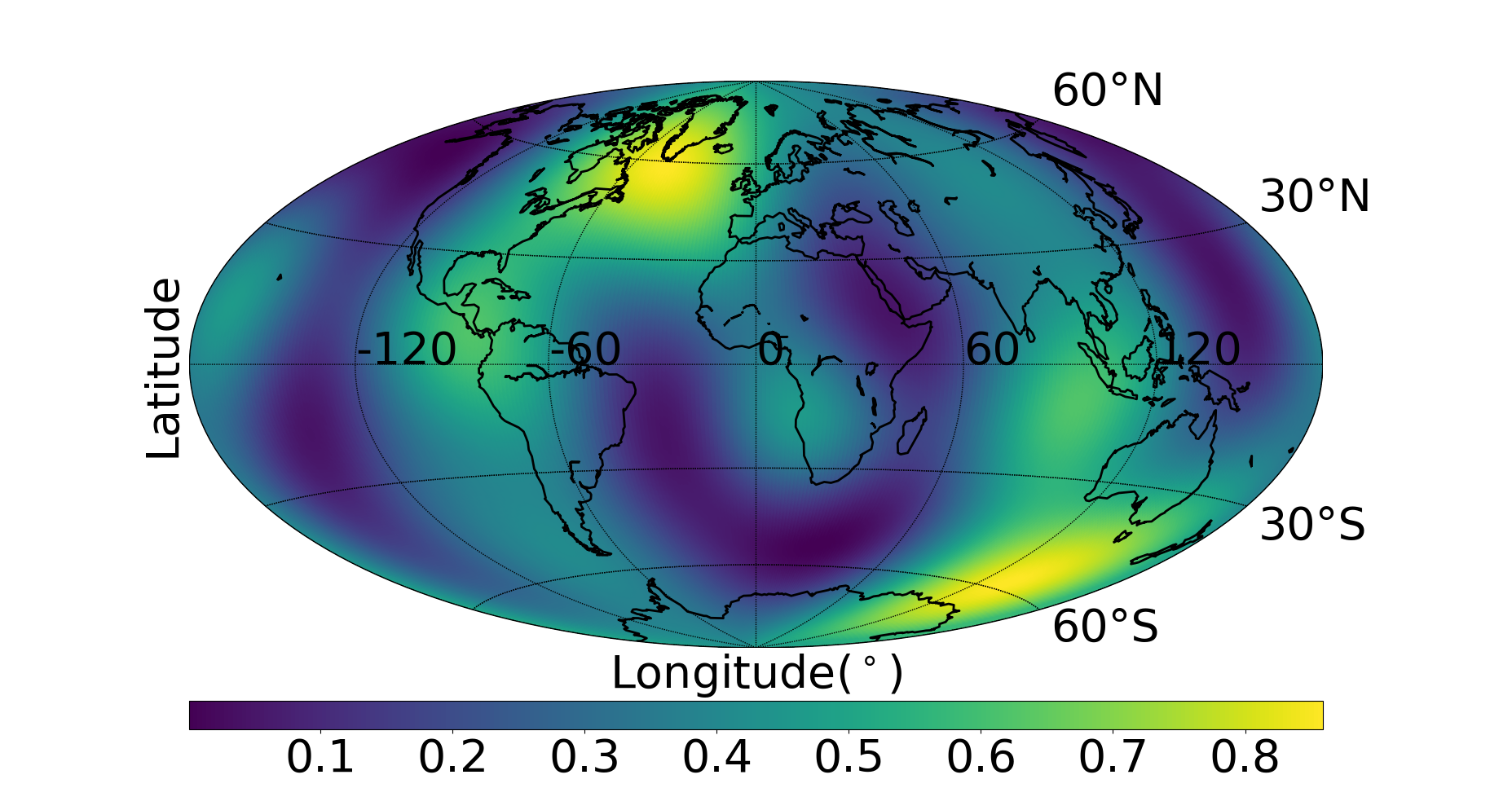}%
        }\quad
        \subfigure[]{
            \label{fig:tap}
            \includegraphics[width=0.48\textwidth]{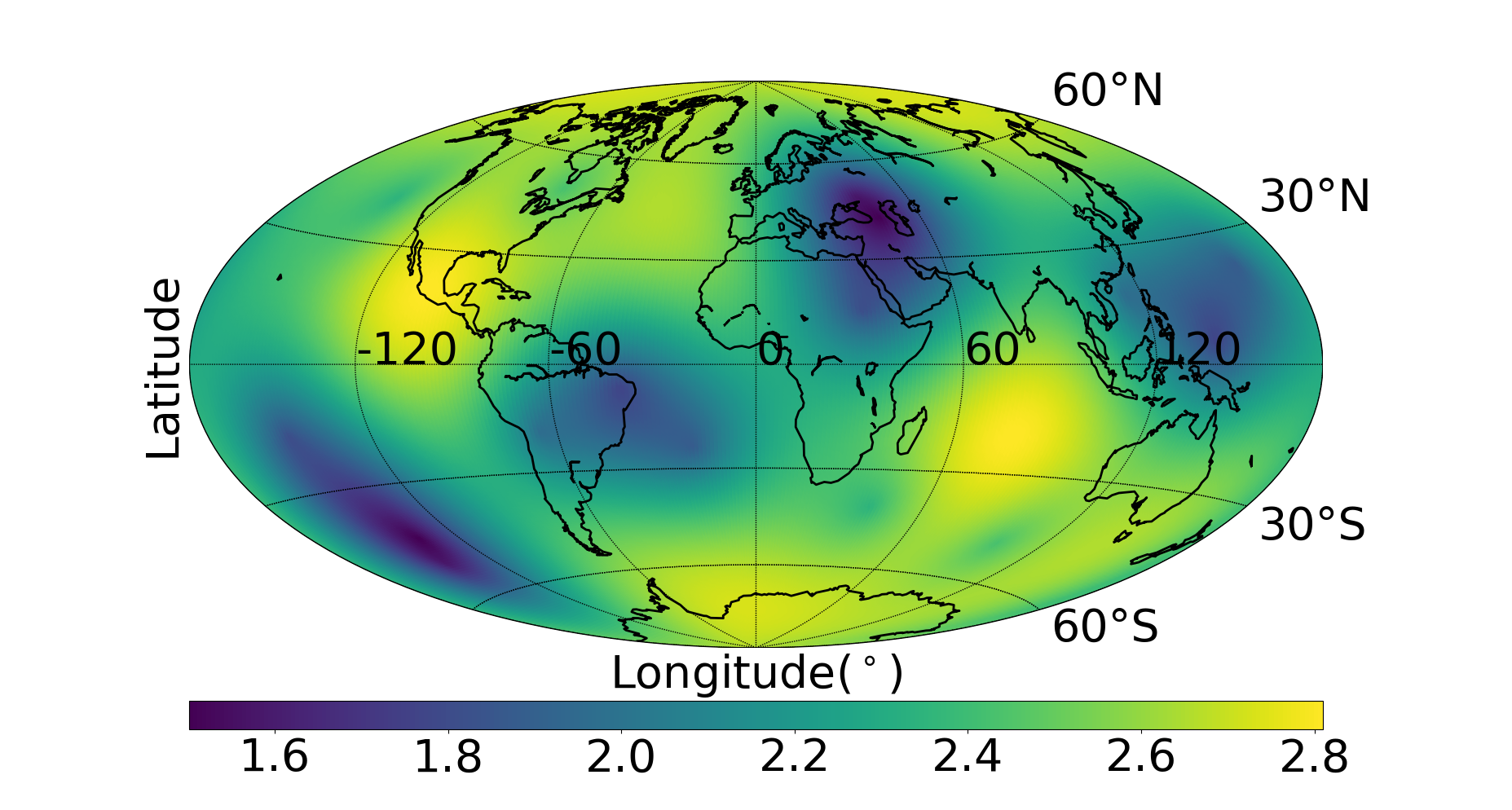}%
        }
    \end{center}
    \caption{The skymap on the left shows the distribution of the value of the denominator in Eq. \ref{eq:reconstruction} for a network of \ac{GW} detectors consisted of \ac{aLIGO} Hanford, \ac{aLIGO} Livingston,
\ac{AdVirgo} and KAGRA, while the skymap on the right shows the distribution of $A = \sqrt{\sum_{j} F_{j+}^2 +  F_{j\times}^2}$ for the same network. 
    \label{fig:domonapcom}} 
\end{figure*}

As indicated in Eq. \ref{eq:reconstruction}, if at a location in the sky, the value of the denominator 
on the right hand side is close to zero, the values on the right will approach infinity 
and become unphysical. This means at such location, the two polarisations of \acp{GW} cannot be recovered. 
As an example, in Fig. \ref{fig:domon}, we show the distribution of this term across 
the sky for a network of \ac{GW} detectors consisted of \ac{aLIGO} Hanford, \ac{aLIGO} Livingston,
\ac{AdVirgo} and KAGRA. 
Clearly, even for a network of four detectors, there are still regions where the reconstruction of the two polarisations is not achievable.
Shown in Fig. \ref{fig:tap} is the value of $A = \sqrt{\sum_{j} F_{j+}^2 +  F_{j\times}^2}$ for the network, where $j$ indicates the $j$th detector.
If such a value is large at a location, sources coming from this direction are more likely to be detectable to the network.
As one may notice from these two plots, the patterns of the distributions do not completely coincide with each other.
In particular, there are regions where one value is high but the other is low or vice versa.
This means that it is possible the \acp{GW} from a source in the sky may be detectable, while the two polarisations and thus the $V$ mode may not be recoverable.
A question that can be raised, therefore, is how observable the $V$ mode will be if we have a detection of \acp{GW} from sources in the sky? 

\begin{figure*}
     \begin{center}
        \subfigure[]{
            \label{fig:waveform}
            \includegraphics[width=0.48\textwidth]{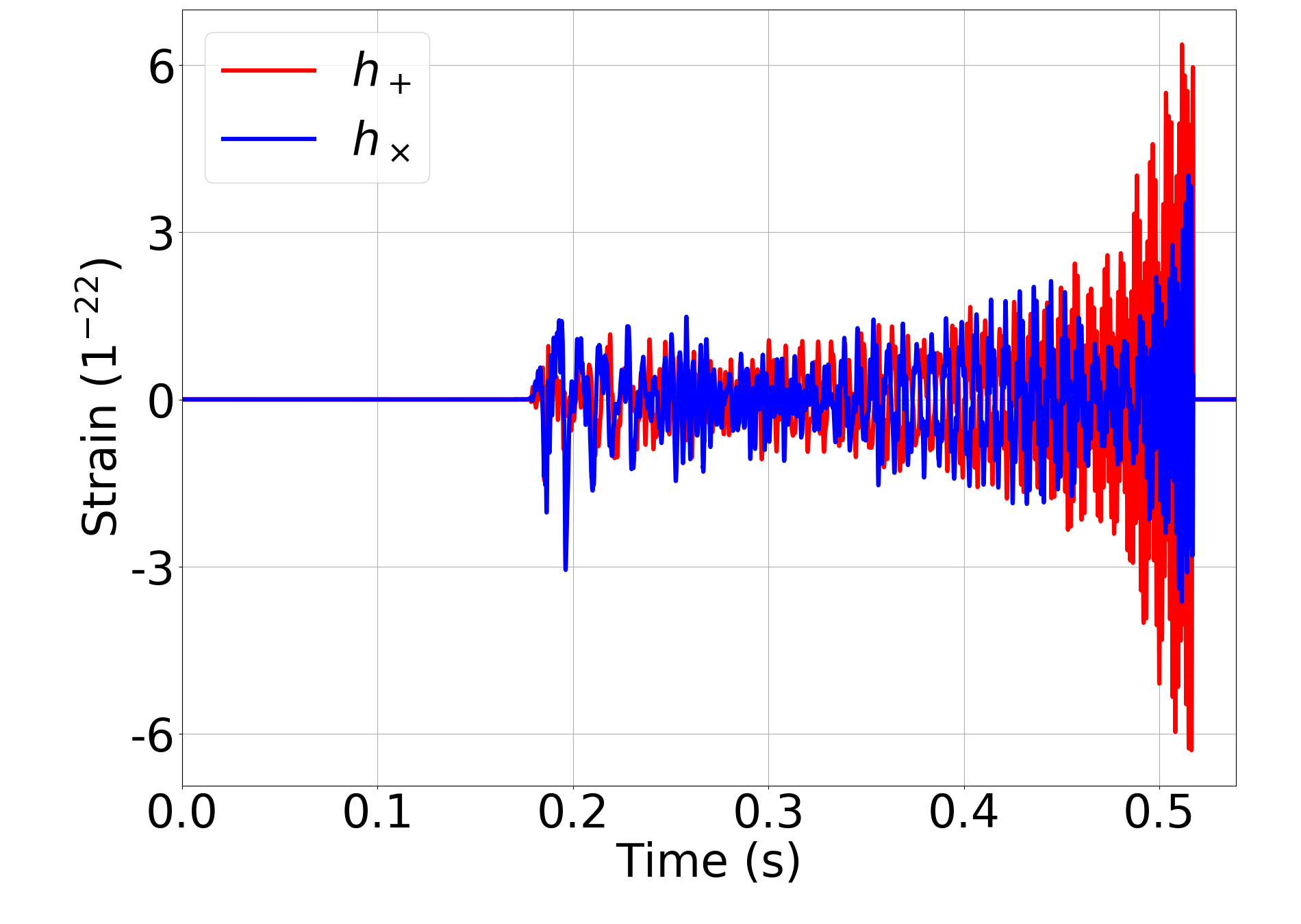}%
        }\quad
        \subfigure[]{
            \label{fig:TFV}
            \includegraphics[width=0.48\textwidth]{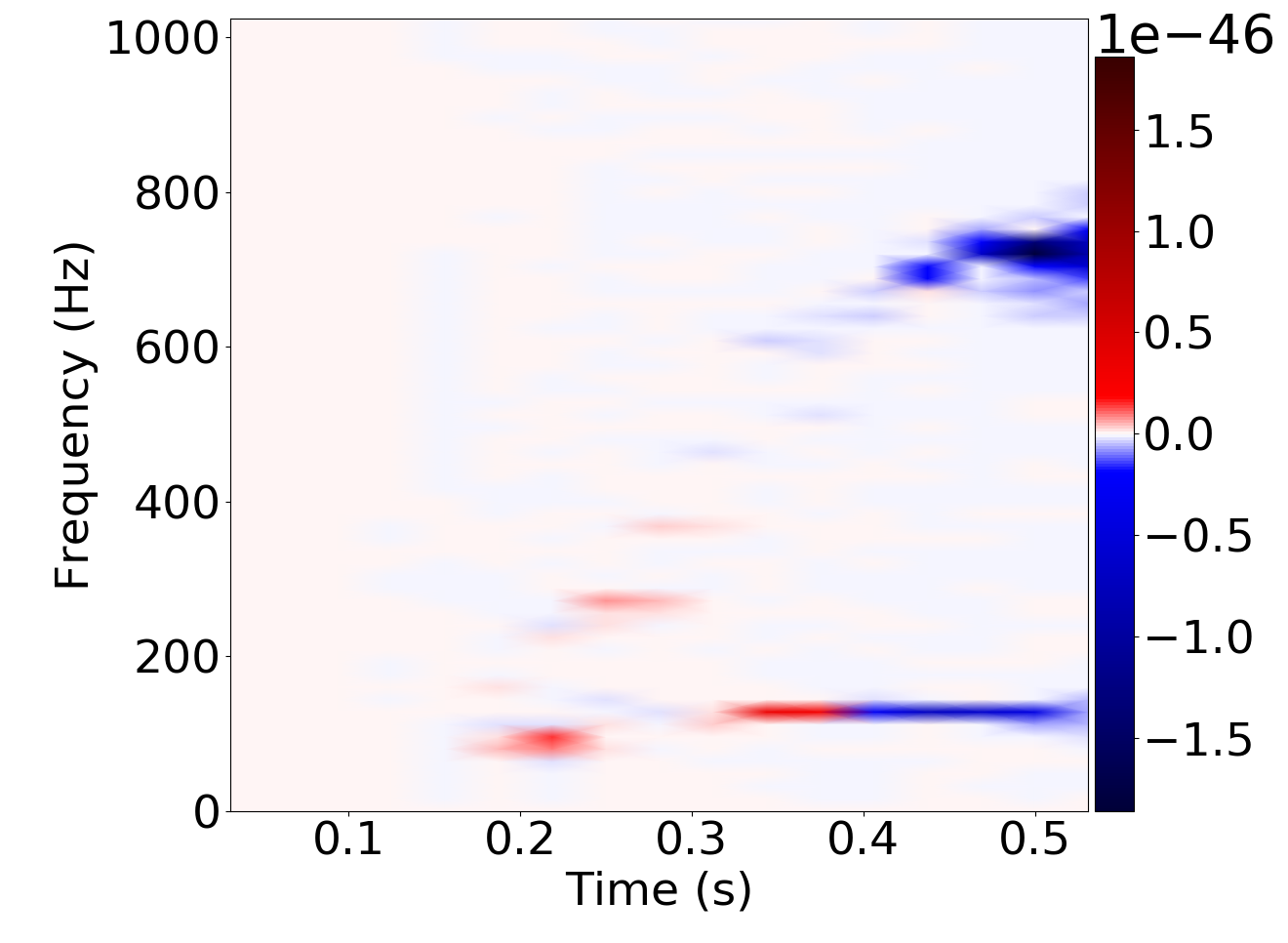}%
        }
    \end{center}
    \caption{Plots showing the waveform SFHx and its $V$ mode.
    The left panel shows the waveform assuming a distance of $10$ kpc from earth.
    The right panel shows the $V$ mode of the waveform at the same distance.
    \label{fig:simpleV2}} 
\end{figure*}

In this section, we try to answer this question by investigating how the values of $\log B_{\text{H}_1 / \text{H}_0}$ 
for a network of the four detectors are distributed. Again, the detectors are the \ac{aLIGO} detectors, \ac{AdVirgo} and KAGRA.
Specifically, we focus on \acp{CCSN} and employ the waveform referred to as SFHx in \cite{kuroda2016new}, shown in Fig. \ref{fig:waveform}. 
The waveform was generated in a 3D full general relativity simulation of \ac{CCSN} assuming a star of $15\text{M}_\odot$ at zero age.
The simulation followed the hydrodynamics of the explosion from the beginning of the collapse for up to $300$ ms after core bounce.
The nuclear equation of state assumed was SFHx \cite{steiner2013core}, 
which is currently considered the best fit model with the observed relation of mass radius of cold neutron stars \cite{steiner2010equation, steiner2013core}.
The waveform is characterised by two distinct features, which are clearly reflected in its $V$ mode shown in Fig. \ref{fig:TFV}.
The first is a power excess increasing from $\sim100$Hz to $\sim800$Hz from after core bounce at $\sim 0.14$s. 
This feature is correlated with the oscillation of the proto-neutron star surface~\cite{muller2013new}.
This feature appears to occur stochastically in the $V$ mode (the seemingly random change of the asymmetry between the right-handed and left handed mode) 
due to Buoyancy-driven proto-neutron star surface oscillation, which also occurs stochastically~\cite{murphy2009model}. 
In addition, a quasi-periodic modulation feature can be seen from $0.14$s to $0.30$s at low frequency ($\sim100$Hz) 
caused by the mass accretion flows striking the proto-neutron star core surface.
The feature is characterised by a dominance of right-handed mode from $\sim 0.14$s to $\sim 0.20$s, which switches to a dominance of left-handed mode 
from $\sim 0.22$s to $\sim 0.32$s. 
Between the dominance of different modes, a quiescent phase where the polarisation of close to zero amplitude is observed.

For the purpose of investigating the detectability of the $V$ mode for sources across the sky, 
we perform simulations at $3$ different distances, 
i.e., $2$ kpc, $5$ kpc and $10$ kpc.
For each distance, we generate $5\times 10^4$ sources distributed in the sky assuming a uniform distribution on right ascension and on the sine of declination.
When computing the value of $\log {B_{\text{H}_1 / \text{H}_0}}$ for an event, we choose $N$ 
(the number of noise only whitened time series in the computation of the $V$ mode) 
to be $2000$ and $F_o$ (the threshold for selecting pixels in the time-frequency domain) to be $0.99$. 
For simplicity, we use simulated Gaussian noise generated using
power spectrum densities of the detectors at their respective design sensitivities. 
Since we are focused on \acp{GW} from \acp{CCSN}, 
we assume that information on the sky locations of the sources and the arrival times of the signals are available from
electromagnetic and/or neutrino observations. 
This means when recovering the $V$ mode of a trigger, 
we use the true values of the $\bold{F}_+$ and $\bold{F}_{\times}$ as well as the arrival times.

\section{Discussion}\label{sec:disuss}
\begin{figure*}
     \begin{center}
        \subfigure[]{
            \label{fig:2kpcskymap}
            \includegraphics[width=0.48\textwidth]{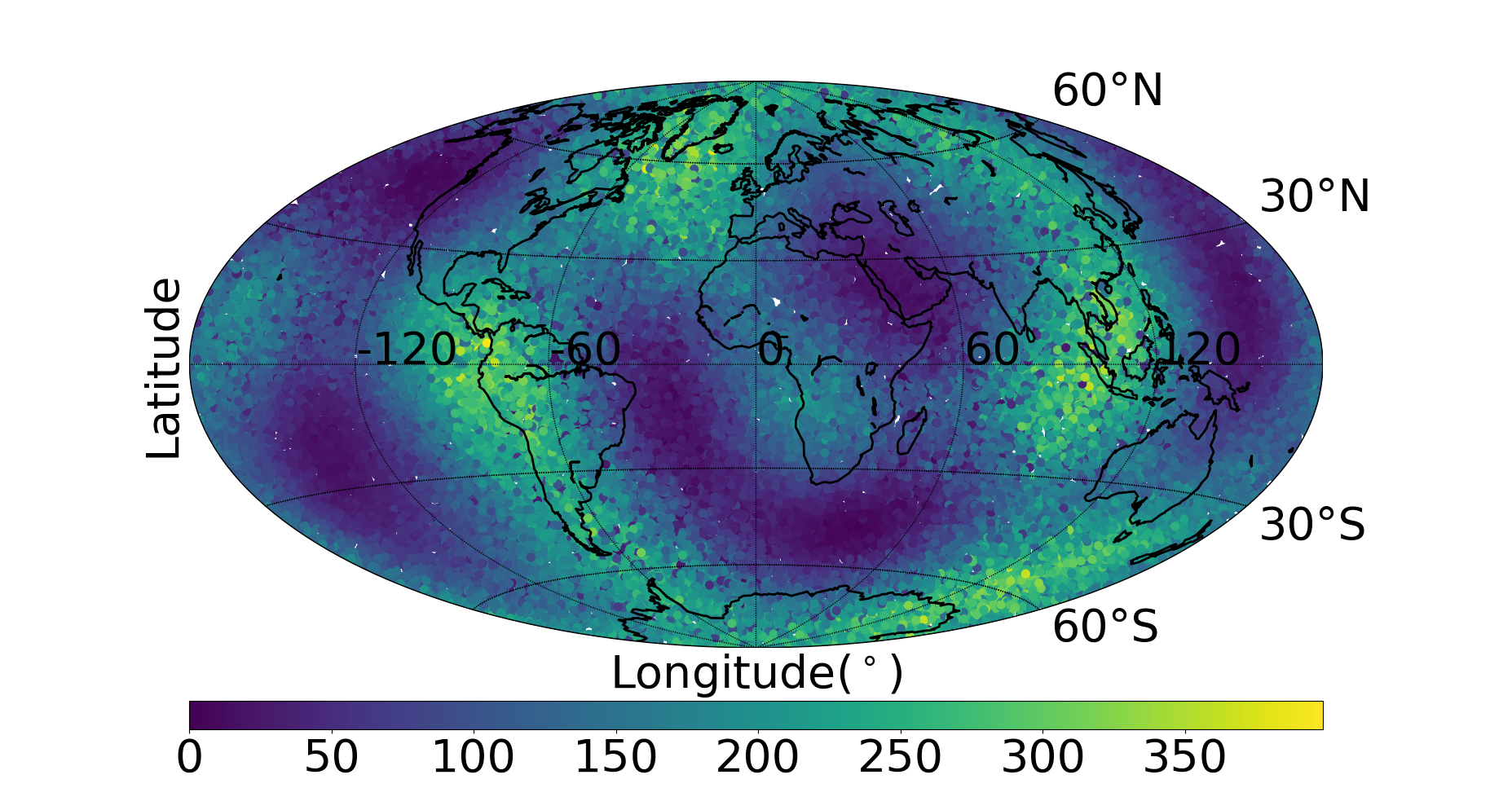}%
        }\quad
        \subfigure[]{
            \label{fig:2kpcdist}
            \includegraphics[width=0.48\textwidth]{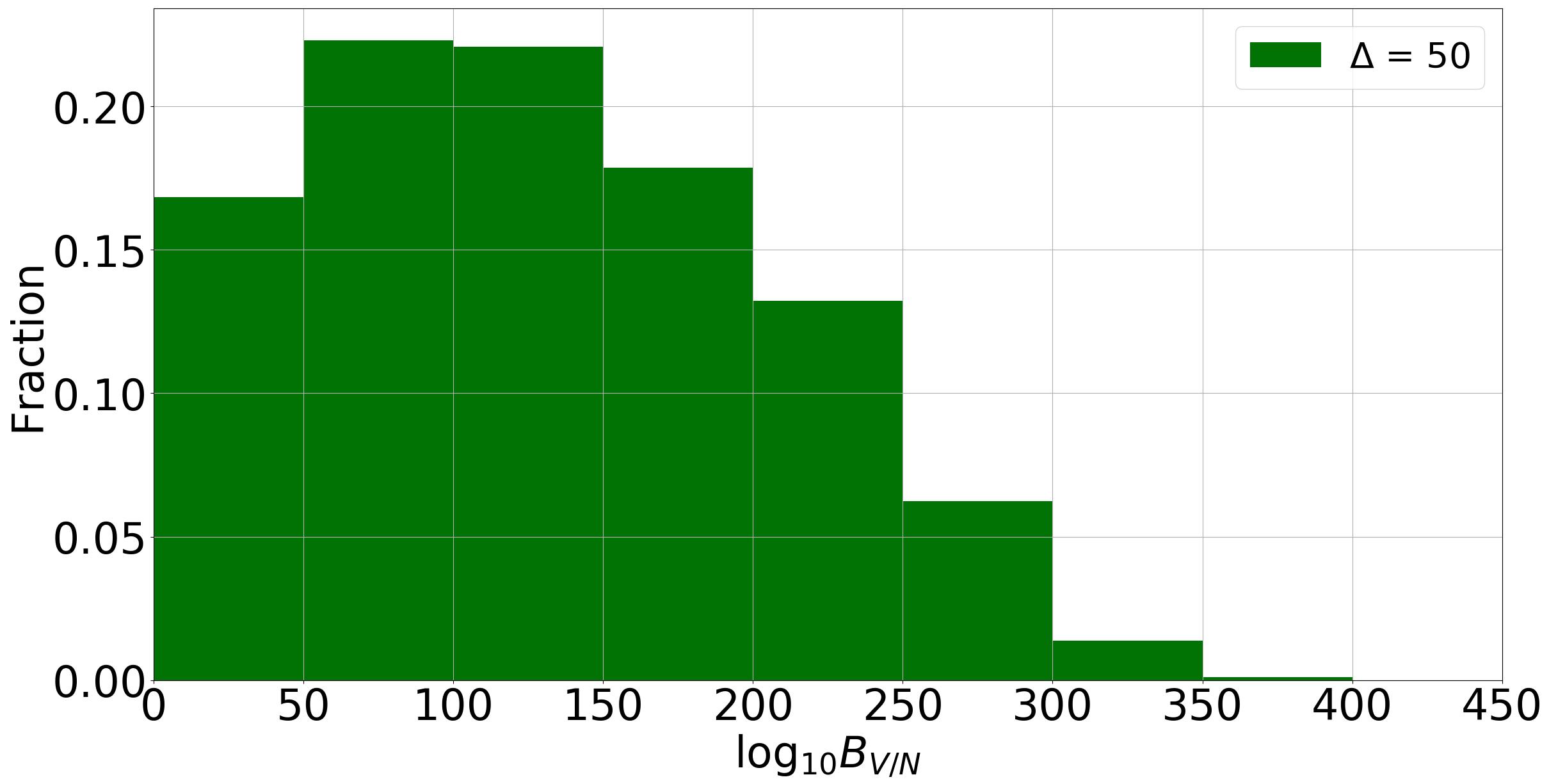}%
        }\quad
        \subfigure[]{
            \label{fig:5kpcskymap}
            \includegraphics[width=0.48\textwidth]{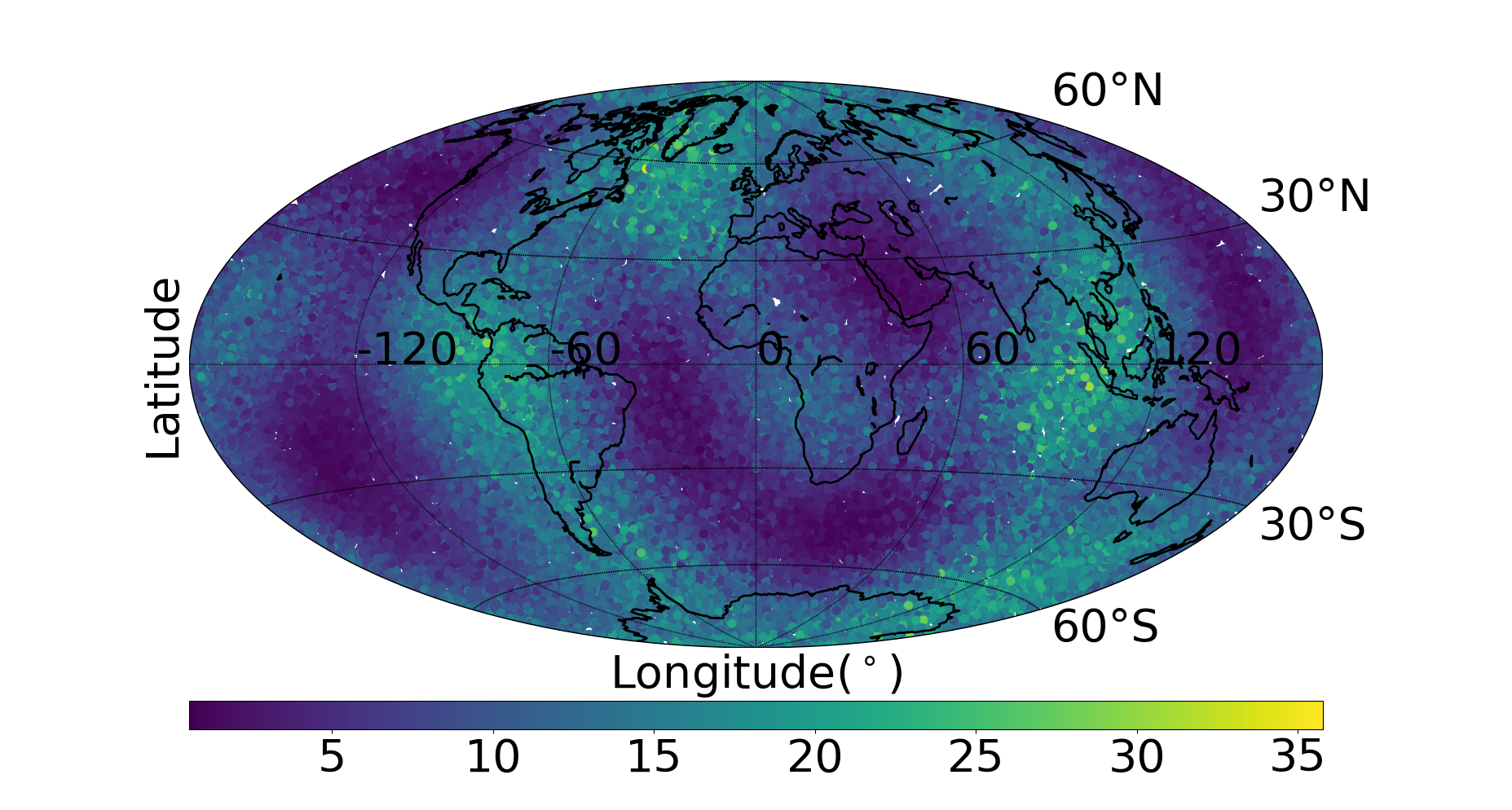}%
        }\quad
        \subfigure[]{
            \label{fig:5kpcdist}
            \includegraphics[width=0.48\textwidth]{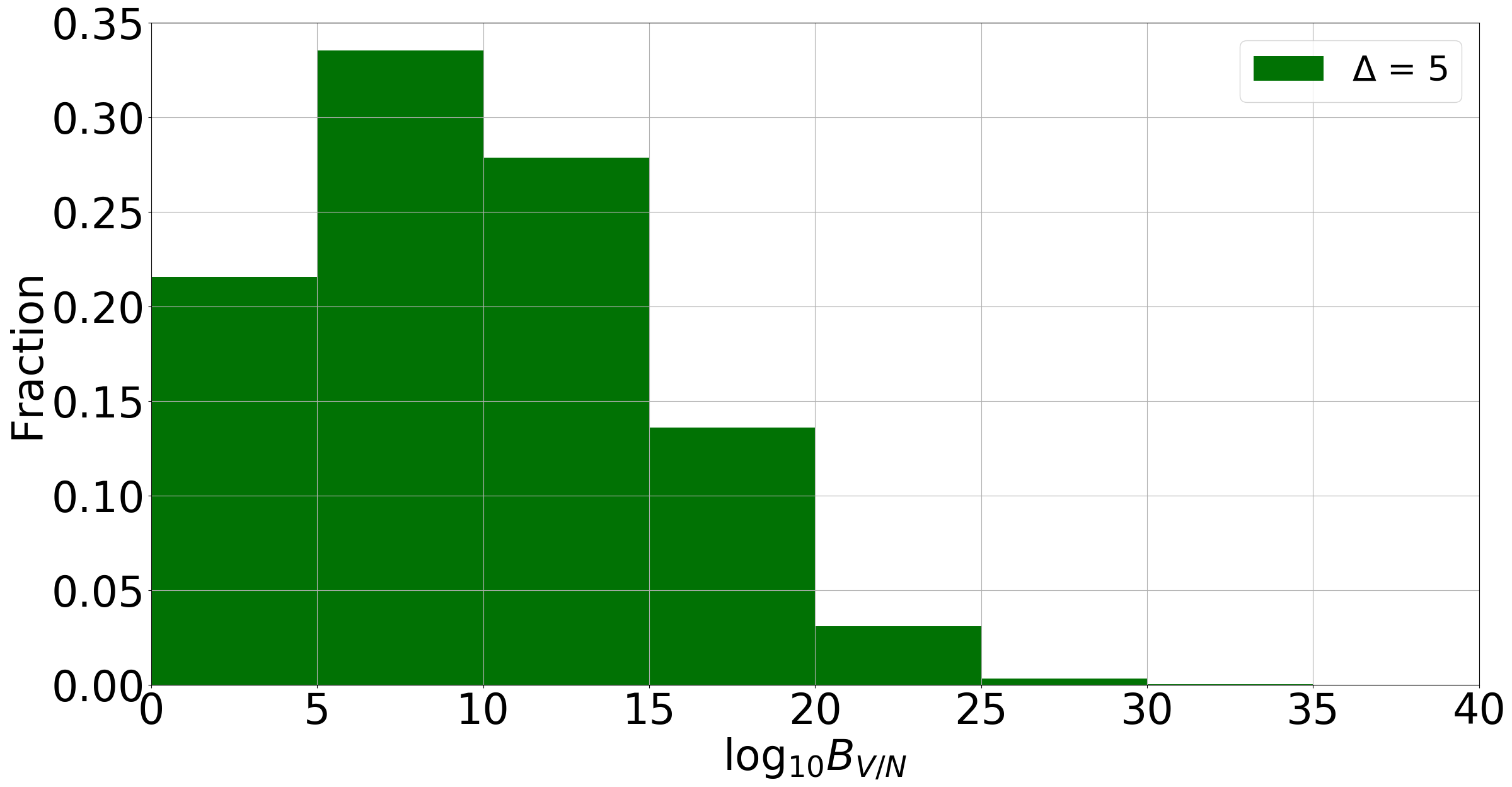}%
        }\quad
        \subfigure[]{
            \label{fig:10skymap}
            \includegraphics[width=0.48\textwidth]{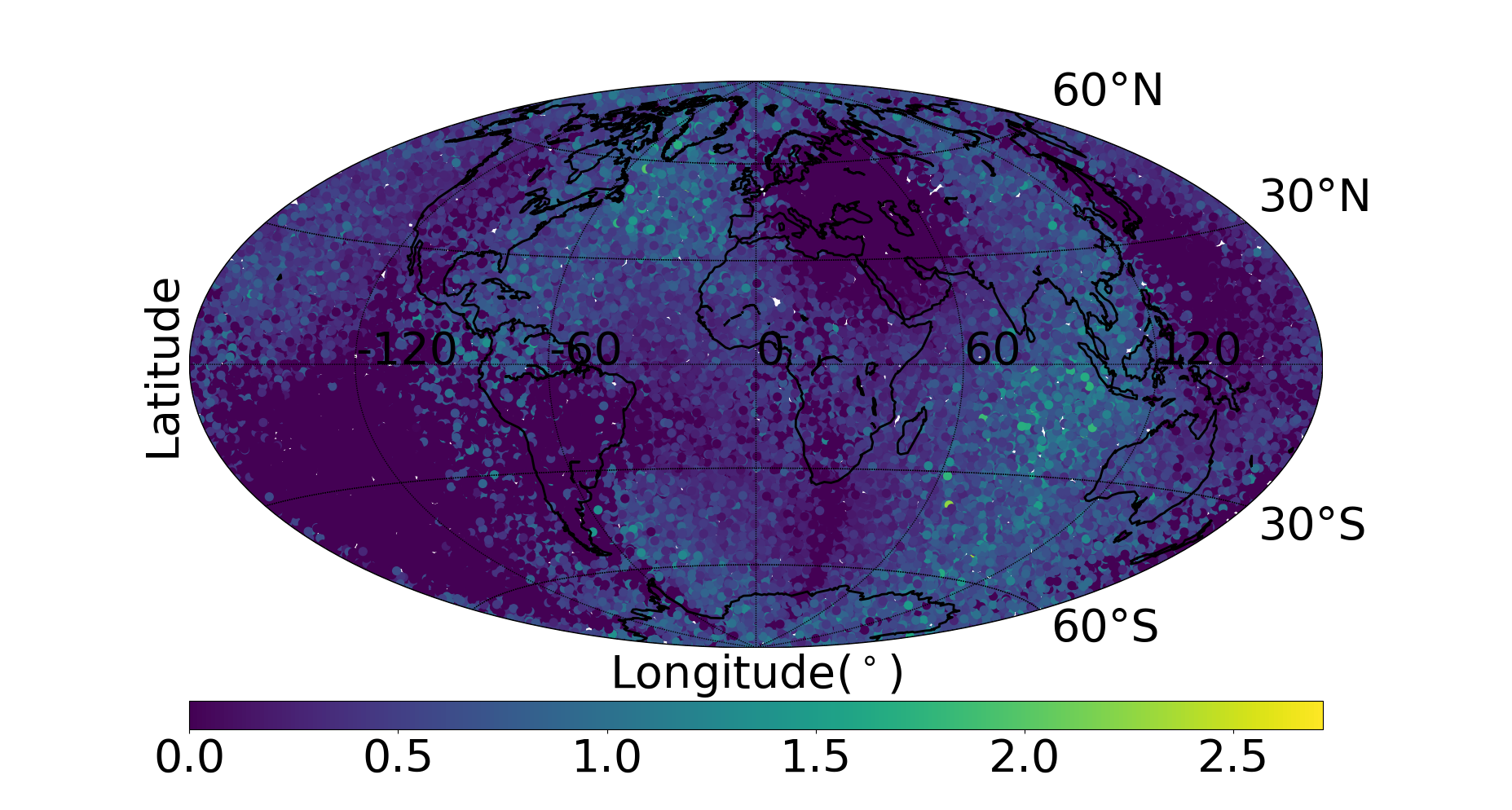}%
        }\quad
        \subfigure[]{
            \label{fig:10kpcdist}
            \includegraphics[width=0.48\textwidth]{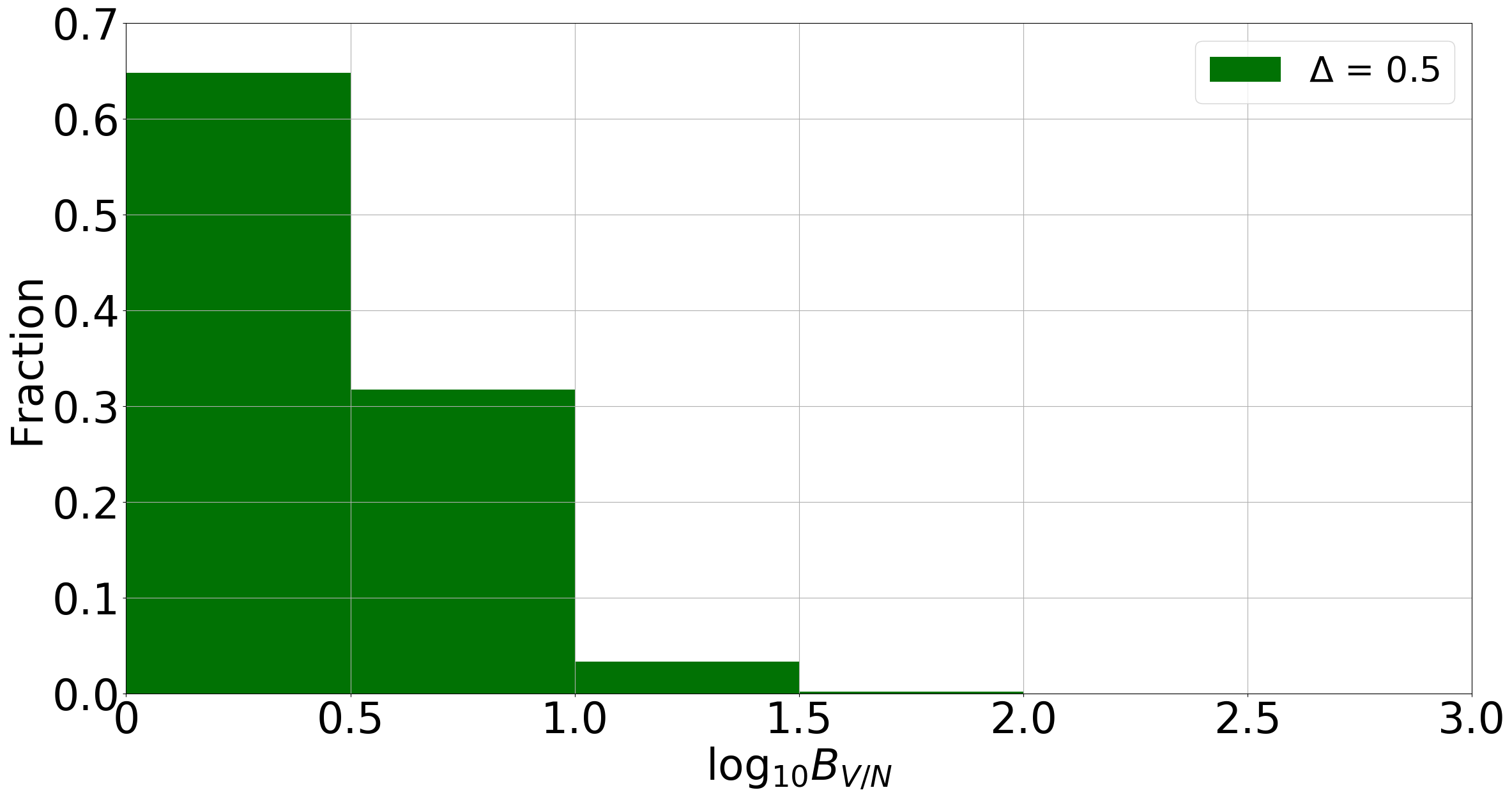}%
        }
    \end{center}
    \caption{The distributions of the values of $\log B_{\text{H}_1 / \text{H}_0}$.
    On the left, the skymaps show how the values of $\log B_{\text{H}_1 / \text{H}_0}$ are distributed across the sky.
    On the right, the plots show the histogram of the values of $\log B_{\text{H}_1 / \text{H}_0}$.
    The value of $\Delta$ in the legends on the right panels are the bin size. 
    From the top to the bottom, the distances are $2$, $5$ and $10$ kpc respectively.
    \label{fig:simulationresults}} 
\end{figure*}

We present the results of the simulations in this section.
Specifically, the results for $2$ kpc, $5$ kpc and $10$ kpc are shown in Fig. \ref{fig:simulationresults}.
It can be seen that there are sky regions where the values of $\log B_{\text{H}_1 / \text{H}_0}$ 
are higher than that at the other regions.
In particular, for sources at $2$ kpc, the distribution of  $\log B_{\text{H}_1 / \text{H}_0}$ forms a pattern that 
is consistent with the pattern shown in Fig. \ref{fig:domon}. This is somewhat expected, 
since the $V$ mode relies on the reconstruction of the two polarisations of \acp{GW}.
However, it is noticeable that regions where the values are the highest in Fig. \ref{fig:domon} are not the highest in Fig. \ref{fig:2kpcskymap},
(For example, one of the highest values in Fig. \ref{fig:domon} occurs at (longitude, latitude) = $(\sim 60^\circ, \sim 60^\circ)$, 
while the largest Bayes factor
for sources at $2$ kpc happens at (longitude, latitude) = $(\sim -100^\circ, \sim 0^\circ)$. )
That is because the amplitudes of the signal are first modulated by the antenna pattern of the detector networks.
This means the distribution of $\log B_{\text{H}_1 / \text{H}_0}$ 
is not only affected by the denominator of the term on the right hand side in Eq. \ref{eq:reconstruction},
but also the combined antenna pattern of the network of \ac{GW} detectors.
For sources at $5$ kpc, the same pattern appears 
with lower values of $\log B_{\text{H}_1 / \text{H}_0}$ as the amplitudes of the signals are inversely proportional to the distance.
The distributions of $\log B_{\text{H}_1 / \text{H}_0}$ for sources at these distances are shown in the right panels of Fig. \ref{fig:simulationresults}.
If a value of $\log B_{\text{H}_1 / \text{H}_0} >= 8$ is required to claim a detection of the $V$ mode, that would be $99.9\%$ and $58.2\%$ of the injections 
for $2$kpc and $5$ kpc respectively.

However, for sources at $10$ kpc, the distribution in the skymap appears to be different.
Since as stated in section \ref{sec:al}, the \ac{SCP} algorithm will be called only when a trigger is identified by the pipeline \ac{CWB}.
For closer distances such as $5$ kpc and $2$ kpc, this may not appear to be a problem, as the majority of the sources are detectable with the \ac{CWB}.
But for $10$ kpc, a noticeable fraction of the sources start to be undetectable and thus resulting in a different distribution of $\log B_{\text{H}_1 / \text{H}_0}$. 
Further, if we again employ the same criterion for $\log B_{\text{H}_1 / \text{H}_0}$
(i.e., $\log B_{\text{H}_1 / \text{H}_0} >=8$), that would mean at $10$ kpc, no $V$ mode is detectable.
We present the fraction of detectable sources and the fraction of sources with $\log B_{\text{H}_1 / \text{H}_0} >=8$ in Table~\ref{table:nr}.
\begin{table}[]
\centering
\begin{threeparttable}
\caption{Numerical results}
\label{table:nr}
\def\arraystretch{2}%
\begin{tabular}{ccc}
\toprule
Distance  & Detectable & $\log B_{\text{H}_1 / \text{H}_0} >=8$ \\
\hline
2 kpc  & 100\%               & 99.9\%                    \\
5 kpc  & 100\%               & 58.2\%                    \\
10 kpc & 69\%                & 0.0\%                     \\
\hline
\hline
\end{tabular}
\begin{tablenotes}
\setlength\labelsep{0pt}
\normalfont{
\item From the left to the right, the first column is the distance for the sources, the second column
mean the fraction of injections that are detectable to \ac{CWB}, and the third column indicates the fraction of the injections
of which the values of $\log B_{\text{H}_1 / \text{H}_0}$ is no less than $8$.}
\end{tablenotes}
\end{threeparttable}
\end{table}

From the results above, it can be seen that there is a relation between the detectability of a signal and that of its $V$ mode.
To show this, we present our results from another perspective.
We plot the values of $\rho$ as a function of $\log B_{\text{H}_1 / \text{H}_0}$, where
$\rho$ is defined by the following equation,
\begin{equation}\label{eq:rho}
 \rho = \sqrt{\frac{\text{E}_{\text{c}}}{\text{J}-1}},
\end{equation}
where $\text{J}$ is the number of detectors in a network, and $\text{E}_{\text{c}}$ is referred to as the coherent energy and is defined in~\cite{klimenko2016method}. 
The value of $\rho$ is one of the most important criteria based on which \ac{CWB} determines if a trigger is identified in the data.
\begin{figure}
     \begin{center}
        \subfigure[]{
            \label{fig:2kpcrhovsbf}
            \includegraphics[width=0.48\textwidth]{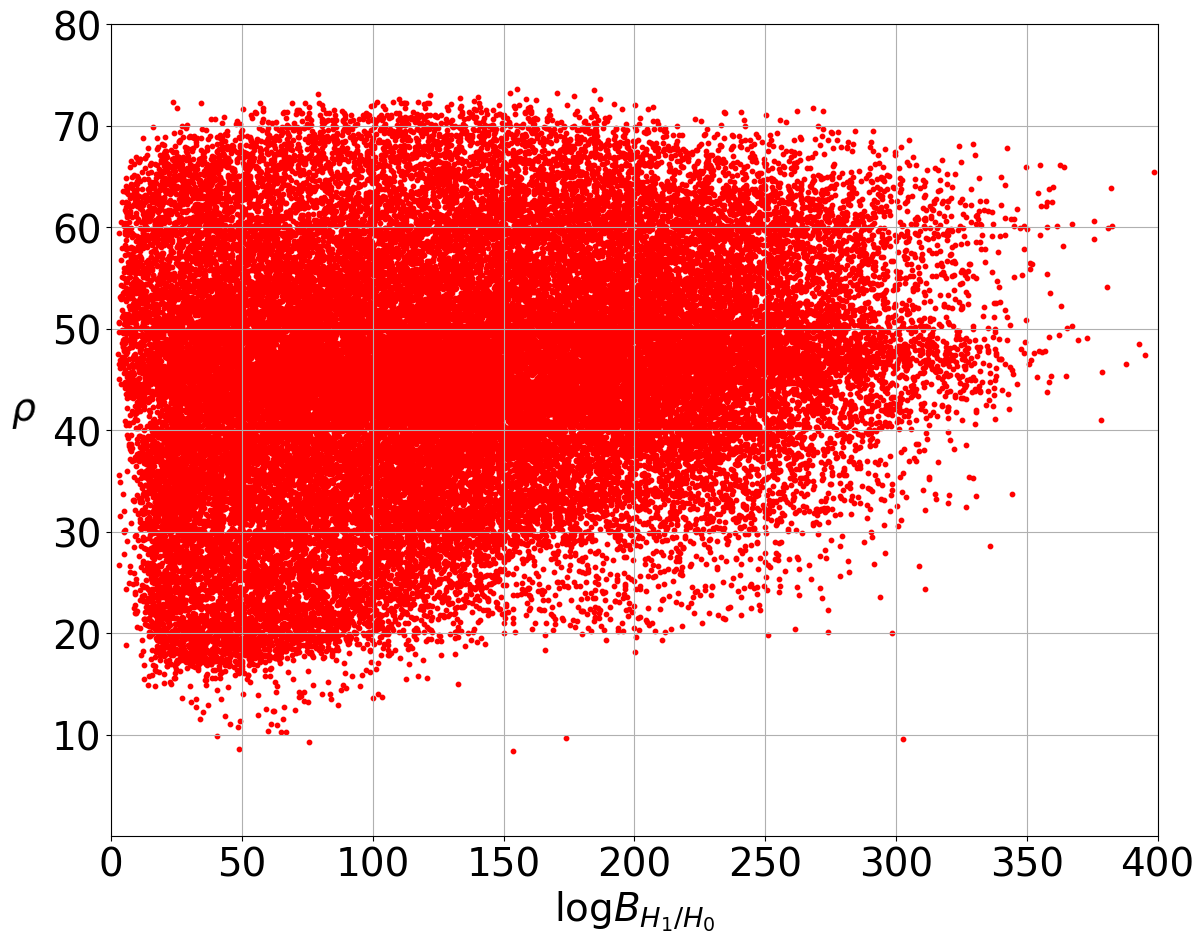}%
        }\quad
        \subfigure[]{
            \label{fig:2kpcrhovsbf}
            \includegraphics[width=0.48\textwidth]{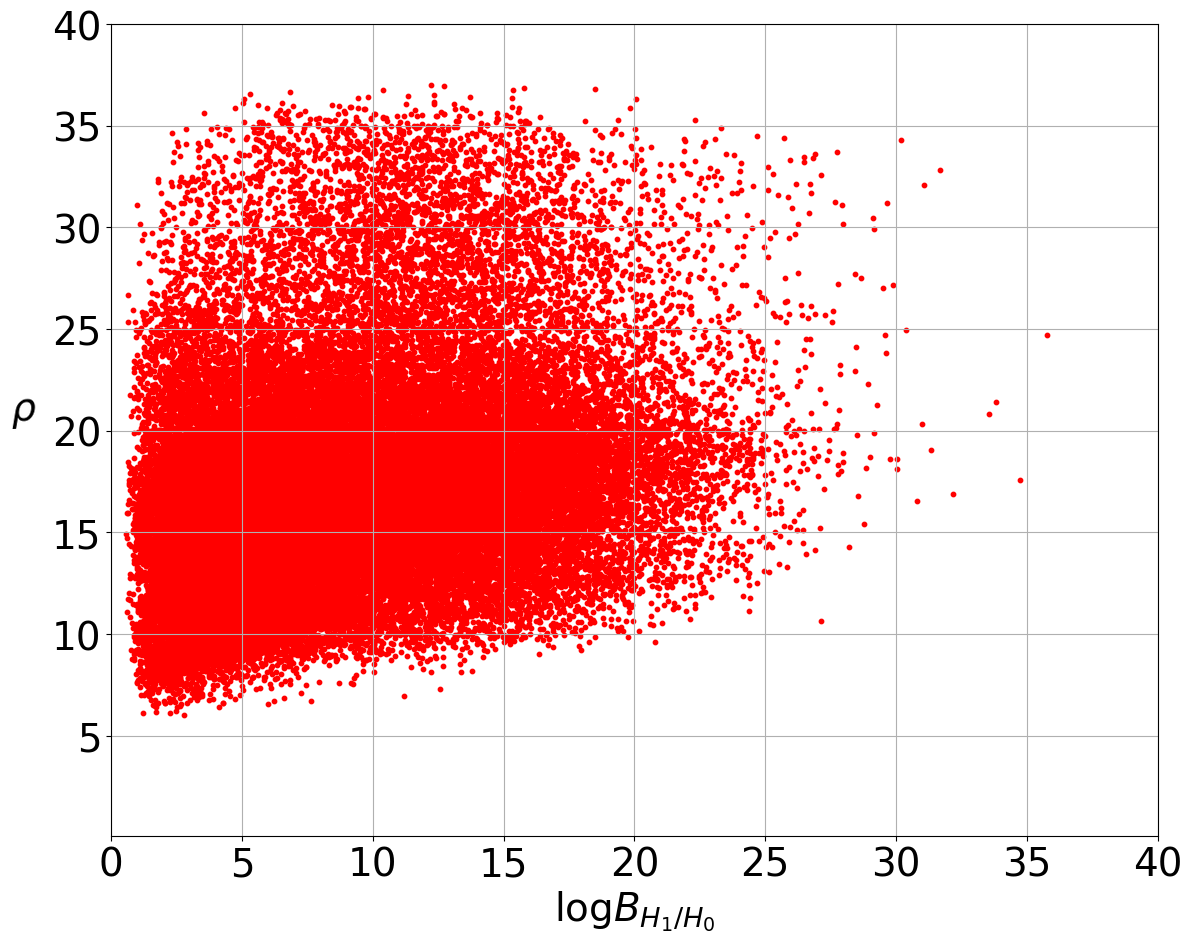}%
        }\quad
        \subfigure[]{
            \label{fig:5kpcrhovsbf}
            \includegraphics[width=0.48\textwidth]{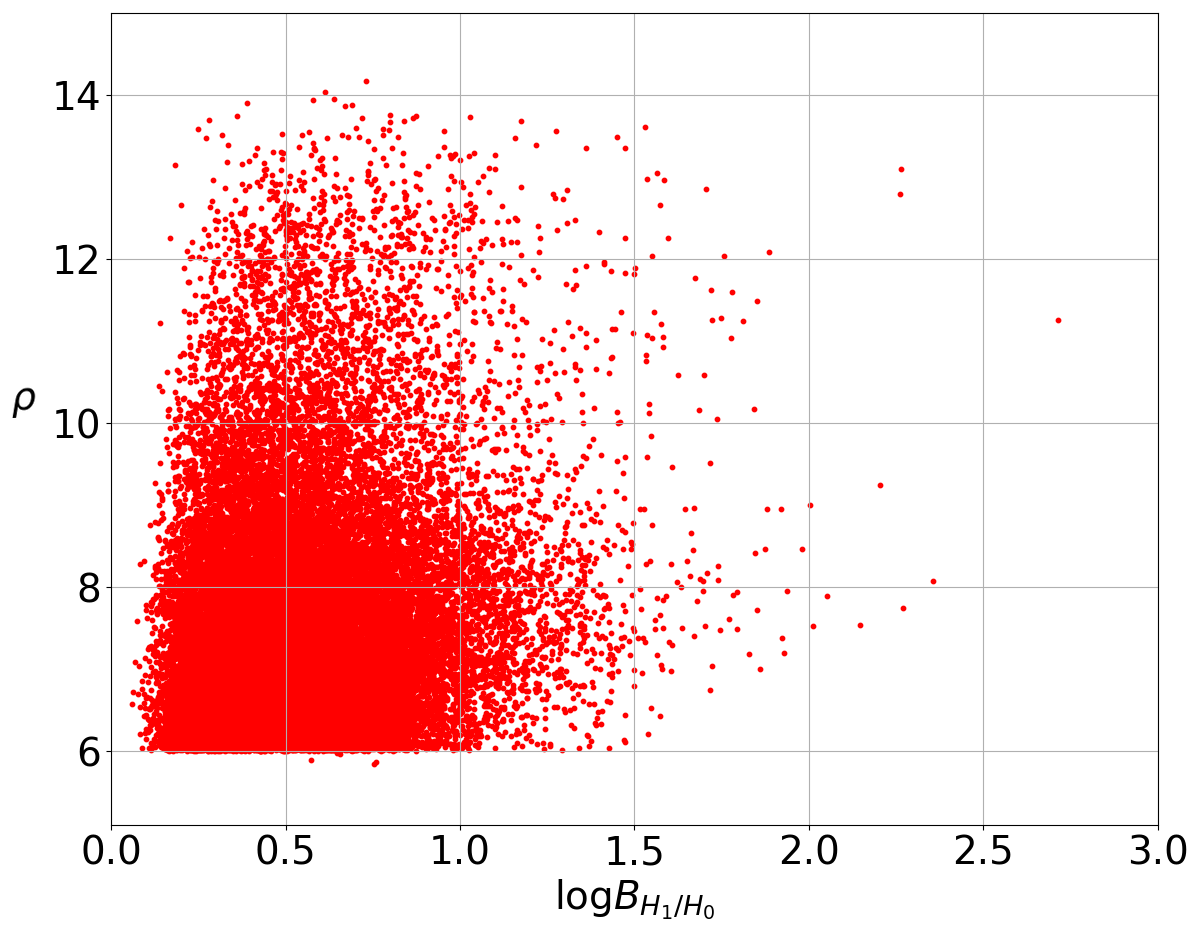}%
        }
    \end{center}
    \caption{Plots showing the values of $\rho$ (see text for definition) as a function of $\log B_{\text{H}_1 / \text{H}_0}$.
    From the top to the bottom, the sources are at $2$, $5$ and $10$ kpc respectively.
    \label{fig:rhovsbf}} 
\end{figure}
In Fig \ref{fig:5kpcrhovsbf}, a low cut-off of $\rho = 6$ can be seen. 
This is caused by the minimal value of $\rho$ required for a trigger.  
Interestingly, for $\log B_{\text{H}_1 / \text{H}_0} > 1.5$, the lowest value of $\rho$ seems to change linearly. 
This proves that the values of $\log B_{\text{H}_1 / \text{H}_0}$ partially depends on the detectability of a signal.

We argue that although the simulations presented in this paper is done using only one waveform, the applicability of the results is not limited. 
If fact, the results can be extrapolated to other waveforms by comparing the $V$ mode of other waveforms with the $V$ mode of the SFHx waveform shown in Fig. \ref{fig:simpleV2}
and the values of $\log B_{\text{H}_1 / \text{H}_0}$ of their $V$ mode using the method laid out in section \ref{sec:al}.
In addition, the method presented in this paper is not limited to the waveform used.

As stated before, the \ac{SCP} algorithm will be called to compute the $V$ mode of a signal 
only if the signal is detectable to \ac{CWB}. 
This requirement has the advantage of increasing the confidence in the event of a detection.
However, it is possible that such a requirement can potentially cause signals to be overlooked that may otherwise be detectable because
the detectability of $V$ modes may not be completely correlated with the detectability of the signals using more traditional methods as shown in~\cite{hayama2018circular}.
For future study, to circumvent this problem, 
we plan to relax such a requirement by expanding the \ac{SCP} algorithm and employing only the $V$ mode as detection statistics.

\section{Conclusion}\label{sec:conclusion}
We have developed an algorithm referred to as the \ac{SCP} algorithm that works with the pipeline \ac{CWB}.
Using the whitened time series from the \ac{CWB} as well as the 
estimates of the sky locations and the arrival times of events or those from electromagnetic and/or neutrino observations,
the algorithm will recover the $V$ mode if a trigger is identified by the \ac{CWB}. 
The Bayes factor is employed to determine whether a $V$ mode signature is significant. 

By using the SFHx waveform as an example and simulating $5\times10^4$ sources for three different distances, 
we investigated the distribution of $\log B_{\text{H}_1 / \text{H}_0}$  for sources across the sky with a network of four \ac{GW} detectors.
We showed that for sources at $2$ kpc and $5$ kpc, the distributions of $\log B_{\text{H}_1 / \text{H}_0}$
are consistent with how regions are distributed where the two polarisations of \acp{GW} are recoverable.
The values of $\log B_{\text{H}_1 / \text{H}_0}$, however, will be partially dependent on the combined antenna pattern of the network.
The distribution of $\log B_{\text{H}_1 / \text{H}_0}$ for sources at $10$ kpc appear to be different as the sources are starting to be undetectable.
Using a criterion of $\log B_{\text{H}_1 / \text{H}_0} >=8 $ for a detection of $V$ mode, we showed that for waveform SFHx, $99.9\%$ and $58.2\%$
of the $V$ mode of sources at $2$ and $5$ kpc are detectable, while for sources at $10$ kpc, no $V$ mode will be detectable.

\section{ACKNOWLEDGEMENTS}
This work is partially supported by JSPS KAKENHI Grant Number JP19K03896, JP17H06364.
We would like to express our gratitudes to Marek Szczepanczyk and Sergey Klimenko for the helpful dicussions of this work.
We are also thankful to Kanda Nobuyuki, Kei Kotake, Takiwaki Tomoya,  Chris Messenger and  Ik Siong Heng, 
for their constructive comments.

\end{document}